\documentstyle[12pt]{article}

\def\beq{\begin{equation}}
\def\eeq{\end{equation}}
\def\barr{\begin{eqnarray}}
\def\earr{\end{eqnarray}}
\def\b{\bigskip}
\def\n{\noindent}
\def\m{\medskip}

\def\nr{nonrelativistic }
\def\r{relativistic }
\def\cs{Chern-Simons }
\def\sd{self-dual }
\def\dag{\dagger}
\begin{document}
\title{Self-Dual Chern-Simons Theories\footnote{Lectures presented at the {\it
$XIII^{th}$ International Symposium ``Field Theory and Mathematical Physics'',
Mt. Sorak, Korea (June-July 1994)}; hep-th/9410065}}

\author{\normalsize{GERALD DUNNE} \\
\normalsize{Department of Physics, University of Connecticut}\\
\normalsize{Storrs, CT 06269, USA}}

\date{}

\maketitle

\begin{abstract}

In these lectures I review classical aspects of the \sd \cs systems which
describe charged scalar fields in $2+1$ dimensions coupled to a gauge field
whose dynamics is provided by a pure \cs Lagrangian. These \sd models have one
realization with \nr dynamics for the scalar fields, and another with \r
dynamics for the scalars. In each model, the energy density may be minimized by
a Bogomol'nyi bound which is saturated by solutions to a set of first-order
self-duality equations. In the \nr case the \sd potential is quartic, the
system possesses a dynamical conformal symmetry, and the \sd solutions are
equivalent to the static zero energy solutions of the equations of motion. The
\nr self-duality equations are integrable and all finite charge solutions may
be found. In the \r case the \sd potential is sixth order and the \sd
Lagrangian may be embedded in a model with an extended supersymmetry. The \sd
potential has a rich structure of degenerate classical minima, and the vacuum
masses generated by the Chern-Simons Higgs mechanism reflect the self-dual
nature of the potential.

\end{abstract}

\section{Introduction : Self-Dual Theories}
\label{sec-intro}

``Self-duality'' is a powerful notion in classical mechanics and classical
field theory, in quantum mechanics and quantum field theory. It refers to
theories in which the interactions have particular forms and special strengths
such that the second order equations of motion (in general, a set of coupled
nonlinear partial differential equations) reduce to first order equations which
are simpler to analyze. The ``self-dual point'', at which the interactions and
coupling strengths take their special self-dual values, corresponds to the
minimization of some functional, often the energy or the action. This gives
self-dual theories crucial {\it physical} significance. For example, the
self-dual Yang-Mills equations have minimum action solutions known as
instantons, the Bogomol'nyi equations of self-dual Yang-Mills-Higgs theory have
minimum energy solutions known as monopoles, and the Abelian Higgs model has
minimum energy self-dual solutions known as vortices. In these lectures, I
discuss a new class of self-dual theories, {\it \sd \cs theories}, which
involve charged scalar fields minimally coupled to gauge fields whose
`dynamics' is provided by a \cs term in $2+1$ dimensions. The physical context
in which such self-dual models arise is that of anyonic quantum field theory.
An interesting novel feature of these \sd \cs theories is that they permit a
realization with either \r or \nr dynamics for the scalar fields. In the \nr
case, the self-dual point corresponds to a quartic scalar potential, with
overall strength determined by the \cs coupling strength. The \nr \sd \cs
equations may be solved completely for all finite charge solutions, and the
solutions exhibit many interesting relations to two dimensional (Euclidean)
integrable models. In the \r case, while the general exact solutions are not
explicitly known, the solutions correspond to topological and nontopological
solitons and vortices, many characteristics of which can be deduced from
algebraic and asymptotic data. These \sd \cs theories also have the property
that, at the self-dual point, they may be embedded into a model with an
extended supersymmetry, a general feature of self-dual theories.

Before introducing the \sd \cs theories, I briefly review some other important
\sd theories, in part as a means of illustrating the general idea of
self-duality, but also because various specific properties of these theories
appear in our analysis of the \sd \cs systems. More details concerning some of
these models can be found in the lectures of Professor C.~Lee on ``Instantons,
Monopoles and Vortices'' from this symposium.

Perhaps the most familiar, and in a certain sense the most fundamental,
self-dual theory is that of four dimensional self-dual Yang-Mills theory. The
Yang-Mills action is
\barr
{\cal S}_{YM}=\int d^4 x \; tr\, \left( F_{\mu\nu} F_{\mu\nu} \right)
\label{yangmills}
\earr
where $F_{\mu\nu}=\partial_\mu A_\nu -\partial_\nu A_\mu +[A_\mu , A_\nu ]$ is
the gauge field curvature. The Euler-Lagrange equations form a complicated set
of coupled nonlinear partial differential equations:
\barr
D_\mu F^{\mu\nu}=0
\label{ymeqs}
\earr
where $D_\mu=\partial_\mu + [A_\mu , \; ]$ is the covariant derivative.
However, in four dimensional Euclidean space the Yang-Mills action
(\ref{yangmills}) is minimized by solutions of the self-dual (or
anti-self-dual) Yang-Mills equations:
\barr
F_{\mu\nu} = \pm \tilde{F}_{\mu\nu}
\label{sdym}
\earr
where $\tilde{F}_{\mu\nu}\equiv  \epsilon_{\mu\nu\rho\sigma}F_{\rho\sigma}/2$
is the dual field strength. Note that the self-dual equations (\ref{sdym}) are
first order equations (in contrast to the second order equations of motion
(\ref{ymeqs})), and their ``instanton'' solutions are known in detail
\cite{rajaraman}. We shall see that the \nr \sd \cs equations have an
interesting connection with these \sd Yang-Mills equations.

Another important class of self-dual equations are the ``Bogomol'nyi
equations''
\barr
D_i \Phi = -\epsilon_{i\,j\,k}F_{j\,k}
\label{bogomolnyi}
\earr
which arise in the theory of magnetic monopoles in $3+1$ dimensional
space-time. These equations arise from a minimization of the static energy
functional of a Yang-Mills-Higgs system in a special parametric limit known as
the BPS limit \cite{prasad,bog}. It is interesting to note that these
Bogomol'nyi equations can be obtained from the (anti-) self-dual Yang-Mills
equations (\ref{sdym}) by a `dimensional reduction' in which all fields are
taken to be independent of $x^4$, and $A_4$ is identified with $\Phi$:
\barr
F_{41}=F_{23} \hskip 1in &\to& \hskip 1in D_1 \Phi = - F_{23} \cr
F_{42}=-F_{13} \hskip 1in &\to& \hskip 1in D_2 \Phi =  F_{13} \cr
F_{43}=F_{12} \hskip 1in &\to& \hskip 1in D_3 \Phi =  - F_{12}
\earr
We shall see that the \nr \sd \cs equations may also be obtained from the \sd
Yang-Mills equations by a dimensional reduction. Furthermore, the \r \sd \cs
equations involve a special algebraic embedding problem (that of embedding
$SU(2)$ into the gauge algebra) which also plays a crucial role in the analysis
of the Bogomol'nyi equations (\ref{bogomolnyi}).

The abelian Higgs model in $2+1$ dimensions is a model of a complex scalar
field $\phi$ interacting with a $U(1)$ gauge field with conventional Maxwell
dynamics. For a special quartic potential, with a particular overall strength,
the static energy functional is minimized by solutions to the following set of
self-duality equations:
\barr
D_j \phi&=&-i\;\epsilon_{j\, k}D_k \phi\cr
F_{12}&=&1-|\phi |^2
\label{abhiggs}
\earr
These self-duality equations have vortex solutions \cite{nielsen,bog,jaffe}
which are important in the phenomenological Landau-Ginzburg theory of
superconductors. The self-duality equations we find in the \sd \cs systems also
arise from minimizing the energy functional in a $2+1$ dimensional theory, and
the resulting \cs self-duality equations have a similar form to the abelian
Higgs model self-duality equations (\ref{abhiggs}).

Yang \cite{yang} proposed an approach to the four dimensional self-dual
Yang-Mills equations (\ref{sdym}) in which they can be viewed as the
consistency conditions for a set of first order differential operators. This
idea is fundamental to the notion of ``integrability'' of systems of
differential equations, a subject with many connections to self-dual theories
\cite{ward1,das}. If the self-dual Yang-Mills equations (\ref{sdym}) are
rewritten in terms of the null coordinates $u=(x^1+ix^2)/\sqrt{2}$ and
$v=(x^3+ix^4)/\sqrt{2}$, they become
\barr
F_{u\, v}& =&0\cr
F_{\bar{u}\,\bar{v}}&=&0\cr
F_{u\,\bar{u}}+F_{v\,\bar{v}}&=&0
\label{sdym2}
\earr
These express the consistency conditions for the first order equations
\barr
\left( D_u - \zeta\,D_{\bar{v}}\right) \psi &=&0\cr
\left( D_v + \zeta\,D_{\bar{u}}\right) \psi &=&0
\label{firstcomp}
\earr
where $\zeta$ is known as a ``spectral parameter''. The first two equations in
(\ref{sdym2}) can be solved locally to give
\barr
A_{u}=H^{-1}\partial_{u} H \hskip 1.5in A_{v}=H^{-1}\partial_{v} H\cr
A_{\bar{u}}=K^{-1}\partial_{\bar{u}} K \hskip 1.5in
A_{\bar{v}}=K^{-1}\partial_{\bar{v}} K
\earr
where $H$ and $K$ are gauge group elements. Then, defining $J=H K^{-1}$, the
third of the self-duality equations in (\ref{sdym2}) becomes
\barr
\partial_{\bar{u}} \left( J^{-1} \partial_{u} J\right)+ \partial_{\bar{v}}
\left( J^{-1} \partial_v J\right)=0
\label{newyang}
\earr
If we now make a dimensional reduction in which the fields are chosen to be
independent of $x^2$ and $x^4$, this equation becomes the two dimensional
equation
\barr
\partial_\mu \left( J^{-1} \partial_\mu J \right) = 0
\label{chimod}
\earr
which is known as the chiral model equation. The chiral model equation will
play a very important role in our analysis of the \nr \sd \cs equations. Also
note that if $J\in SU(N)$ and $J$ is further restricted to satisfy the
condition $J^2={\bf 1}$, then (\ref{chimod}) is the equation of motion for the
$CP^{N-1}$ model \cite{rajaraman,zak1}.

The final class of models which we shall recall in this introduction are known
as Toda theories. The original Toda system \cite{toda} described the
displacements of a line of masses joined by springs with an exponential spring
tension. The equations of motion for the Toda lattice are
\barr
\ddot{y_i} = - C_{i\, j} e^{\displaystyle y_j}
\label{originaltoda}
\earr
where the matrix $C_{i\, j}$ is the tridiagonal discrete approximation to the
second derivative, and can be chosen for periodic or open boundary conditions.
This system is classically integrable in the limit of an infinite number of
masses, in the sense that it possesses an infinite number of conserved
quantities in involution. The Toda lattice system also has a deep algebraic
structure due to the fact that the matrix $C_{i\,j}$ in (\ref{originaltoda}) is
the Cartan matrix of the Lie algebra $SU(N)$ (or its affine extension). Indeed,
this relationship allows one to extend the original Toda system to a Toda
lattice based on other Lie algebras \cite{kostant,leznov1,perelomov}.

The Toda system generalizes still further, to an integrable set of {\it
partial} differential equations
\barr
\nabla^2 {y_i} = - C_{i\, j} e^{\displaystyle y_j}
\label{todapde}
\earr
which is not only integrable, but also {\it solvable}, in the sense that the
solution may be written in terms of $2r$ arbitrary functions, where $r$ is the
rank of the classical Lie algebra whose Cartan matrix appears in
(\ref{todapde}) \cite{kostant,leznov1}. For $SU(2)$ the classical Toda system
reduces to the nonlinear Liouville equation
\barr
\nabla^2 y = - 2 e^{\displaystyle y}
\label{origliouville}
\earr
which was solved by Liouville \cite{liouville}. Both the Liouville and Toda
equations, together with their solutions, appear prominently in the analysis of
the \nr \sd \cs models. Moreover, the Toda equations also arise from the
Bogomol'nyi equations (\ref{bogomolnyi}) when one looks for spherically
symmetric monopole solutions \cite{leznov2}. This reduction involves an
algebraic embedding problem very similar to one that appears in the treatment
of the \r \sd \cs models.

The \sd \cs theories discussed in these lectures describe charged scalar fields
in $2+1$ dimensional space-time, minimally coupled to a gauge field whose
dynamics is given by a \cs Lagrangian rather than the conventional Maxwell (or
Yang-Mills) Lagrangian. The possibility of describing gauge theories with a \cs
term rather than with a Yang-Mills term is particular to odd-dimensional
space-time, and $2+1$ dimensions is special in the sense that the derivative
part of the \cs Lagrangian is {\it quadratic} in the gauge fields. To conclude
this introduction, I briefly review some of the important properties
\cite{deser1,witten,dunne1} of the \cs Lagrange density:
\barr
{\cal L}_{CS}&=& \epsilon^{\mu\nu\rho} tr \left ( \partial_\mu A_\nu A_\rho
+{2\over 3} A_\mu A_\nu A_\rho \right )
\label{cslag}
\earr
The gauge field $A_\mu$ takes values in a finite dimensional representation of
the gauge Lie algebra ${\cal G}$. The totally antisymmetric $\epsilon$-symbol
$\epsilon^{\mu\nu\rho}$ is normalized with $\epsilon^{012}=1$. The
Euler-Lagrange equations of motion derived from this Lagrange density are
simply
\barr
F_{\mu\nu}=0
\label{flat}
\earr
which follows directly from the fact that
\barr
{\delta {\cal L}_{CS} \over \delta A_\mu } = \epsilon^{\mu\nu\rho} F_{\nu\rho}
\label{var}
\earr
The equations of motion (\ref{flat}) are gauge covariant under the gauge
transformation
\barr
A_\mu \to A_\mu^g\equiv g^{-1}A_\mu g +g^{-1}\partial_\mu g
\label{gt}
\earr
and so the Lagrange density (\ref{cslag}) defines a sensible gauge theory  even
though (\ref{cslag}) is not invariant under the gauge transformation
(\ref{gt}). Indeed, under a gauge transformation ${\cal L}_{CS}$ transforms as
\barr
{\cal L}_{CS}\left( A\right) \to {\cal L}_{CS}\left(A\right) -
\epsilon^{\mu\nu\rho}\partial_\mu tr \left( \partial_\nu g\,g^{-1}\,
A_\rho\right) -{1\over 3}\epsilon^{\mu\nu\rho} tr \left( g^{-1}\partial_\mu g
g^{-1}\partial_\nu g g^{-1}\partial_\rho g \right)
\label{change}
\earr
For an abelian \cs theory, the final term in (\ref{change}) vanishes and the
change in ${\cal L}_{CS}$ is a total space-time derivative. Hence the {\it
action} $S=\int d^3x {\cal L}_{CS}$ is gauge invariant. However, for a
nonabelian \cs theory the final term in (\ref{change}) is proportional to the
winding number of the group element $g$, and the action changes by a constant.
To ensure that $exp(i\,S)$ remains invariant, the \cs Lagrange density
(\ref{cslag}) must be multiplied by a dimensionless coupling parameter $\kappa$
which assumes quantized values \cite{deser1,witten}
\barr
\kappa = {integer\over 4\pi}
\label{quantization}
\earr
The \cs term describes a {\it topological} gauge field theory \cite{witten} in
the sense that there is no explicit dependence on the space-time metric. This
follows because the Lagrange density (\ref{cslag}) can be written directly as a
3-form $tr (AdA+A^3)$. This fact implies that if the \cs Lagrange density
${\cal L}_{CS}$ is coupled to other fields, then it will not contribute to the
energy momentum tensor. This may also be understood by noting that ${\cal
L}_{CS}$ is first order in space-time derivatives
\barr
{\cal L}_{CS}=\epsilon^{i\,j} tr \left( A_i \dot{A}_j\right) + tr \left( A_0
F_{1\,2}\right)
\label{firstorder}
\earr
The time derivative part of ${\cal L}_{CS}$ contributes to the canonical
structure of the theory, the $A_0$ part contributes to the Gauss law
constraint, and there is no contribution to the Hamiltonian. It is very
important that ${\cal L}_{CS}$ is first order in space-time derivatives,
because in the \sd \cs theories discussed in these lectures the self-duality
equations (which should be first order) involve the \cs equations of motion
directly.

\section{Nonrelativistic SDCS Theories}
\label{sec-nr}

\n{2.1 {\it Nonrelativistic Self-Dual Chern-Simons Equations}}

\b

The \nr \sd \cs system is a model in $2+1$ dimensional space-time describing
charged scalar fields $\Psi$ with \nr dynamics, minimally coupled to gauge
fields $A_\mu$ with \cs dynamics \cite{jackiw1,grossman,dunne2}. The Lagrange
density for such a system is:
\barr
{\cal L}&=&-\kappa {\cal L}_{CS} +i\, tr \left ( \Psi^\dag D_0 \Psi \right )
-{1\over 2m} tr \left ( ( D_i \Psi )^\dag D_i \Psi \right ) +{1\over 4 m
\kappa} tr \left ( [ \Psi , \Psi^\dag ]^2 \right ) \cr &&
\label{nrlag}
\earr
where ${\cal L}_{CS}$ is the \cs Lagrange density (\ref{cslag}). I have chosen
to work with adjoint coupling of the scalar and gauge fields (for other
couplings of matter and gauge fields see \cite{dunne1}), with the covariant
derivative in (\ref{nrlag}) being $D_\mu \Psi \equiv \partial_\mu \Psi + [
A_\mu , \Psi ]$. The scalar fields $\Psi$ and the gauge fields $A_\mu$ take
values in the same representation of the gauge Lie algebra ${\cal G}$. In these
lectures, ${\cal G}$ will usually be taken to be $SU(N)$, but much of the
formal structure generalizes straightforwardly to other gauge algebras. The
parameter $\kappa$ appearing in (\ref{nrlag}) is the dimensionless \cs coupling
constant, while $m$ denotes the scalar field mass. Notice that the scalar field
potential appearing in (\ref{nrlag}) has a particular {\it quartic} form, with
an overall scale depending on both $m$ and $\kappa$. This form of the potential
is fixed by the condition of self-duality, as shown below.

The Euler-Lagrange equations of motion that follow from the \nr \sd \cs
Lagrange density (\ref{nrlag}) are:
\barr
i D_0 \Psi&=&-{1\over 2m}\vec{D}^2\Psi -{1\over 2m\kappa}[~[\Psi,\Psi^{\dag}],
\Psi]
\label{matter}
\earr
\barr
F_{\mu \nu} &= &-{i\over 2\kappa} \epsilon_{\mu \nu \rho} J^\rho\
\label{gauge}
\earr
where $F_{\mu \nu}=\partial_\mu A_\nu -\partial_\nu A_\mu +[A_\mu,A_\nu]$ is
the gauge curvature, and $J^\rho$ is the covariantly conserved ($D_\mu J^\mu
=0$) nonrelativistic matter current
\barr
J^0 &=&[\Psi,\Psi^{\dag}]\nonumber\\
J^i&=&-{i\over 2m}\left([\Psi^{\dag},D_i \Psi] - [(D_i \Psi)^{\dag},
\Psi]\right)
\label{current}
\earr
In addition there is an abelian current $Q^\rho$
\barr
Q^0 &=&tr \left ( \Psi \Psi^\dag \right )\cr
Q^i &=& -{i\over 2m} tr \left( \Psi^{\dag}D_i \Psi - (D_i \Psi)^{\dag}
\Psi\right)
\label{abcurrent}
\earr
which is ordinarily conserved ($\partial_\mu Q^\mu =0$). The matter field
equation of motion (\ref{matter}) is referred to as the {\it gauged planar
nonlinear Schr\"odinger equation} \cite{jackiw1}. The study of the nonlinear
Schr\"odinger equation in $2+1$-dimensional space-time is partly motivated by
the significance of the $1+1$-dimensional nonlinear Schr\"odinger equation.
Here we consider a {\it gauged} nonlinear Schr\"odinger equation in which we
have not only the nonlinear potential term for the matter fields, but also we
have a coupling of the matter fields to the gauge fields. The gauge equation of
motion (\ref{gauge}) relates the matter and gauge fields via a \cs coupling.
Notice that even though the \cs Lagrange density ${\cal L}_{CS}$ is not
strictly invariant under a gauge transformation, the equations of motion
(\ref{matter},\ref{gauge}) are gauge covariant.

The Hamiltonian density corresponding to the Lagrange density (\ref{nrlag}) is
\barr
{\cal E}&=&{1\over 2m} tr \left ( ( D_i \Psi )^\dag D_i \Psi \right ) - {1\over
4 m \kappa} tr \left ( [ \Psi , \Psi^\dag ]^2 \right )
\label{nrham}
\earr
where we recall that the \cs term ${\cal L}_{CS}$ does not contribute to the
energy density since it is first order in space-time derivatives. The energy
density (\ref{nrham}) is supplemented by the Gauss law constraint
\barr
J^0=- 2 i \kappa F_{12}
\label{gauss}
\earr
which is the ${\rm zero}^{\rm th}$ component of the gauge equations of motion
(\ref{gauge}). To obtain a Bogomol'nyi - style lower bound for the energy
density we employ the following useful identity:
\barr
tr \left ( ( D_i \Psi )^\dag D_i \Psi \right ) = tr \left ( ( D_- \Psi )^\dag
D_- \Psi \right ) - i\, tr \left ( \Psi^\dag \, [ F_{12} , \Psi ] \right )
\label{bochner}
\earr
where $D_\pm \equiv D_1 \pm i\, D_2$.

Using this identity in (\ref{nrham}), together with the Gauss law constraint
(\ref{gauss}) which relates the ``magnetic field'' $F_{12}$ to the \nr matter
charge density $[ \Psi , \Psi^\dag ]$, we see that the energy density can be
written as
\barr
{\cal E}= {1\over 2m} tr \left ( ( D_- \Psi )^\dag D_- \Psi \right )
\label{nrbog}
\earr
This energy density is therefore minimized by solutions of the {\it \nr \sd \cs
equations} :
\barr
D_- \Psi&=&0
\label{nrsd1}
\earr
\barr
\partial_+ A_- -\partial_- A_+ +[A_+,A_-] &=& {1\over\kappa}~[\Psi,
\Psi^{\dag}]\
\label{nrsd2}
\earr

Notice that these self-duality equations are indeed first-order in derivatives
of the fields, in contrast to the gauged nonlinear Schr\"odinger equation
(\ref{matter}) which is second order.

Since the \sd solutions minimize the Hamiltonian density, they provide {\it
static} solutions to the Euler-Lagrange equations of motion
(\ref{matter},\ref{gauge}). Alternatively, one can see this directly from
inspection of the static equations of motion. Note that if $D_{-}\Psi=0$, then
the currents take the simple form
\barr
J^{+}&\equiv&J^1 +i J^2\nonumber\\
&=&-{i\over 2m}[\Psi^{\dag}, D_+ \Psi]
\label{nrsdcurrent}
\earr
The gauge equation of motion (\ref{gauge}) then implies that $A_0={i\over
4m\kappa} [\Psi, \Psi^{\dag}]$. Together with the identity
\barr
\vec{D}^2 \Psi&\equiv &D_+ D_- \Psi + i [F_{12}, \Psi]\nonumber\\
&=&D_+ D_- \Psi - {1\over 2\kappa} [ [\Psi, \Psi^{\dag}], \Psi]
\label{ }
\earr
this reduces the matter equation of motion (\ref{matter}) to
\beq
i\partial_0 \Psi= -{1\over 2m} D_+ D_- \Psi
\label{sdschrodinger}
\eeq
the RHS of which which vanishes for self-dual solutions.

In fact, owing to a remarkable dynamical $SO(2,1)$ symmetry of the
\nr \sd \cs model (\ref{nrlag}), it is possible to show that the self-dual
solutions (\ref{nrsd1},\ref{nrsd2}) saturate {\it all} static solutions of the
equations of motion \cite{dunne2,jackiw2}. For the Abelian models, this fact
has recently been formulated in terms of a Kaluza-Klein reduction of a \r
symmetry \cite{horvathy}.

An important property of the \nr \sd \cs equations (\ref{nrsd1},\ref{nrsd2}) is
that they can be obtained by dimensional reduction from the four dimensional
\sd Yang-Mills equations for a nonAbelian gauge theory. The signature $(2,2)$
SDYM equations are
\barr
F_{12}=F_{34} \hskip 1in F_{13}=F_{24} \hskip 1in F_{14}=-F_{23}
\label{22sdym}
\earr
Taking all fields to be independent of $x^3$ and $x^4$, these reduce to
\barr
F_{12}=[A_3,A_4] \hskip .75in D_1A_3=D_2A_4 \hskip .75in D_1A_4 =-D_2A_3
\label{red}
\earr
which are just the \nr \sd \cs equations (\ref{nrsd1},\ref{nrsd2}) with the
identification $\Psi=\sqrt{\kappa}(A_3-iA_4)$. These dimensionally reduced
self-dual Yang-Mills equations have been studied in the mathematical literature
\cite{hitchin,donaldson}.

\b

\n{2.2 {\it Algebraic Ansatze and Toda Theories}}

\b

Before classifying the general solutions to the \nr self-dual Chern-Simons
equations, it is instructive to consider certain special cases in which
simplifying algebraic {\it Ans\"atze} for the fields reduce
(\ref{nrsd1},\ref{nrsd2}) to familiar integrable nonlinear equations. Note that
since we are considering {\it static} fields, the self-duality equations have
the appearance of equations of motion in two dimensional Euclidean space.

First, choose the fields to have the following Lie algebra decomposition
\beq
A_i = \sum_{a=1}^{r} A_i^a H_a \hskip 1in \Psi = \sum_{a=1}^{r}
\psi^a E_a
\label{ans1}
\eeq
Here, $H_a$ refers to the Cartan subalgebra generators and $E_a$ to the simple
root step operator generators of the gauge Lie algebra, normalized according to
a Chevalley basis (for ease of presentation we consider only simply-laced
algebras here) \cite{humphreys} :
\barr
[H_a , H_b ] &=&0\cr
[E_a , E_{-b} ] &=& \delta_{a\,b} H_a \cr
[H_a , E_{\pm b} ] &=& \pm C_{a\, b} E_{\pm b}\cr\cr
tr \left ( E_a E_{-b} \right ) &=&\delta_{a\, b}\cr
tr \left ( H_a H_b \right ) &=& C_{a\, b}\cr
tr  \left( H_a E_{\pm b} \right) &=&0
\label{chevalley}
\earr
The indices $a$ and $b$ run over $1 \dots r$, where $r$ is the rank of the
gauge algebra ${\cal G}$. The $r\times r$ matrix $C_{a\, b}$ is the Cartan
matrix of ${\cal G}$, which expresses the inner products of the simple roots
$\vec{\alpha}^{(a)}$:
\barr
C_{a\, b} = {2 \vec{\alpha}^{(a)} \cdot \vec{\alpha}^{(b)} \over
|\vec{\alpha}^{(b)}|^2}
\label{cartan}
\earr
For $SU(N)$, the classical Cartan matrix is the $(N-1)\times (N-1)$ symmetric
tridiagonal matrix (familiar from the theory of numerical analysis):
\barr
C=\left(\matrix{2&-1&0&\dots&&0\cr -1&2&-1&0&\cr
0&-1&2&-1&0\cr \vdots&&&&&\vdots\cr
0&\dots&&0&-1&2}\right)
\label{suncartan}
\earr
With the {\it ansatze} ({\ref{ans1}) for the fields, the first of the \nr \sd
\cs equations, $D_- \Psi =0$, reduces to the set of equations
\barr
\partial_- {\rm log} \psi_a = - \sum_{b=1}^r C_{a\, b} A_-^b
\earr
When combined with its adjoint, and with the other \nr \sd \cs equation, we
find the classical Toda equations
\beq
\nabla ^2 {\rm log} \rho_a = -{1\over\kappa} \sum_{b=1}^r C_{a\, b}\; \rho_b
\label{toda}
\eeq
where $\rho_a \equiv |\psi^a |^2$. For $SU(2)$, $r=1$ and the Cartan matrix is
just the single number 2, so the Toda equations (\ref{toda}) reduce to
the Liouville equation
\barr
\nabla ^2 {\rm log}\rho = -{2\over\kappa} \rho
\label{liouville}
\earr
which Liouville showed to be integrable and indeed "solvable" \cite{liouville}
- in the sense that the general real solution can be expressed in terms of a
single holomorphic function $f=f(x^-)$:
\beq
\rho={\kappa} \nabla ^2 {\rm log} \left(1+f(x^-) \bar{f}(x^+)\right)\
\label{liou-sol}
\eeq
Kostant \cite{kostant}, and Leznov and Saveliev \cite{leznov1} have shown that
the classical Toda equations (\ref{toda}) are similarly integrable (and indeed
solvable), with the general real solutions for $\rho_a$ being expressible in
terms of $r$ arbitrary holomorphic functions, where $r$ is the rank of the
algebra. For $SU(N)$ it is possible to adapt the Kostant-Leznov-Saveliev
solutions to a simple form reminiscent of the Liouville solution
(\ref{liou-sol}):
\beq
\rho_a={\kappa} \nabla ^2 {\rm log~det}\left(M_{a}^{\dag}(x^+)
M_{a}(x^-)\right)\
\label{suntoda}
\eeq
where $M_a$ is the $N{\times} a$ {\it rectangular} matrix $M_a =
(u~\partial_-u~\partial_-^2 u~\dots \partial_-^{a-1} u)$, with $u$ being
an $N$-component column vector containing $N-1$ arbitrary holomorphic functions
$f_1(x^-)$, $f_2(x^-)$, $\dots$, $f_{N-1}(x^-)$:
\beq
u=\left(\matrix{1\cr f_1(x^-)\cr f_2(x^-) \cr \vdots\cr
f_{N-1}(x^-)\cr}\right)\
\label{uvec}
\eeq

An alternative, extended, {\it ansatz} for the fields involves the matter field
choice
\beq
\Psi = \sum_{a} \psi^{a} E_{a} +\psi^M E_{-M}\
\label{ans2}
\eeq
where $E_{-M}$ is the step operator corresponding to minus the maximal root.
With the gauge field still as in (\ref{ans1}), the \nr \sd \cs
equations then combine to give the {\it affine} Toda equations
\beq
\nabla ^2 {\rm log} \rho_a = -{1\over\kappa} \sum_{b=1}^{r+1} \tilde{C}_{a b}
\rho_b\
\label{affinetoda}
\eeq
where $\tilde{C}$ is the $(r+1)\times (r+1)$ affine Cartan matrix. These affine
Toda equations are also known to be integrable
\cite{kostant,leznov1,perelomov}, although it is not possible to write simple
convergent expressions such as (\ref{suntoda}) for the solutions.

There is another useful way to understand these various algebraic reductions of
the \nr \sd \cs equations. In two dimensional space we can express the gauge
field as
\barr
A_- &=& G^{-1} \partial_- G\cr
A_+&\equiv &-A_-^\dag
\label{yang}
\earr
where $G$ is an element of the complexification of the gauge group
\cite{yang,mickelsson}. $G$ can be decomposed as
\barr
G = H\; U
\earr
where $H$ is hermitean and $U$ is unitary. Note that only with $H={\bf 1}$ does
(\ref{yang}) correspond to a pure gauge. Gauge transformations on $A_\pm$
correspond to different choices of the unitary factor $U$. In general, the
field strength corresponding to (\ref{yang}) is
\barr
F_{+-}= - U^\dag \left ( H \partial_+ \left ( H^{-2}\partial_- H^2 \right )
H^{-1} \right ) U
\label{yangcurvature}
\earr
With the gauge field represented as in (\ref{yang}), the solution to the
self-duality equation $D_- \Psi=0$ is trivially:
\barr
\Psi = G^{-1} \Psi_0 (x^+) G
\earr
for {\it any} $\Psi_0 (x^+)$. Inserting this solution in the other self-duality
equation yields the gauge invariant equation for $H$:
\barr
\partial_+ \left (H^{-2} \partial_- H^2 \right ) = \Psi_0^\dag H^{-2} \Psi_0
H^2 - H^{-2} \Psi_0 H^2 \Psi_0^\dag
\label{general}
\earr
Thus far, no special choices have been made and equation (\ref{general}) is
still completely general. Now, if we choose to write $H^2$ as
\barr
H^2 = e^{\displaystyle \Phi}
\label{ans3}
\earr
where $\Phi$ is restricted to the Cartan subalgebra, then (\ref{general})
simplifies to
\barr
\partial_+ \partial_- \Phi = \Psi_0^\dag e^{\displaystyle -\Phi} \Psi_0
e^{\displaystyle \Phi} - e^{\displaystyle -\Phi} \Psi_0 e^{\displaystyle \Phi}
\Psi_0^\dag
\label{yangtoda}
\earr
These equations follow as equations of motion from the two-dimensional
Euclidean Lagrange density
\barr
{\cal L} = tr \left ( \partial_\mu \Phi \partial_\mu \Phi + \Psi_0^\dag
e^{\displaystyle -\Phi} \Psi_0 e^{\displaystyle \Phi} \right )
\label{todalag}
\earr
If $\Psi_0(x^+)$ is now chosen to be the {\it constant} field $\Psi_0=\sum_{a}
E_a$ then this Lagrangian (\ref{todalag}) becomes that of the {\it classical}
$SU(N)$ Toda theory, while if $\Psi_0(x^+)$ is chosen to be the constant field
$\Psi_0=\sum_{a} E_a +E_{-M}$ then it becomes that of the {\it affine} $SU(N)$
Toda theory. With these choices for $\Psi_0$ the self-duality equation
(\ref{yangtoda}) reduces to the classical or affine Toda system, respectively.

\b

\n{2.3 {\it Chiral Model, Unitons and General Solutions}}

\b

Having considered some special cases in which the \nr \sd \cs equations reduce
to well-known integrable equations in two-dimensional Euclidean space, we now
consider the question of finding the most general solutions. The key to the
possibility of finding all solutions lies in the fact that there exists a
special gauge transformation $g$ which converts the two equations
(\ref{nrsd1},\ref{nrsd2}) into a {\it single} equation
\beq
\partial_- \chi = [\chi^{\dag}, \chi ]\
\label{single}
\eeq
where $\chi$ is the gauge transformed matter field $\chi = \sqrt{{1\over
\kappa}} g \Psi g^{-1}$. The existence of such a gauge transformation $g^{-1}$
follows from the following zero-curvature formulation of the self-dual
Chern-Simons equations \cite{dunne2,dunne3}. Define
\beq
{\cal{A}}_+ \equiv A_+ - \sqrt{{1\over \kappa}} \Psi \hskip 1in {\cal{A}}_-
\equiv A_- + \sqrt{{1\over \kappa}} \Psi^{\dag}\
\label{curl}
\eeq
Then the \nr self-dual Chern-Simons equations ({\ref{nrsd1},\ref{nrsd2})
together imply that the gauge curvature associated with ${\cal{A}}_\pm$
vanishes:
\barr
{\cal F}_{+-}&=&\partial_+{\cal{A}}_- -\partial_-{\cal{A}}_+ +[{\cal{A}}_+,
{\cal{A}}_-]\cr
&=&0
\label{zerocurv}
\earr
Therefore, at least locally, one can write $\cal{A}_\pm$ as a pure gauge
\beq
{\cal{A}}_\pm  = g^{-1} \partial_\pm g \
\label{pure}
\eeq
for some $g$ in the gauge group. Gauge transforming the \nr self-dual
Chern-Simons equations (\ref{nrsd1},\ref{nrsd2}) with this group element
$g^{-1}$ leads to the single equation (\ref{single}).

Equation (\ref{single}) can be converted into the chiral model equation
\beq
\partial_+ (h^{-1} \partial_- h) + \partial_- (h^{-1} \partial_+ h) = 0\
\label{chiral}
\eeq
upon defining
\barr
\chi\equiv{1\over 2} h^{-1}\partial_+ h
\label{chifield}
\earr
for some $h$ in the gauge group (the fact that it is possible to write $\chi$
in this manner is a consequence of (\ref{single})). Given any solution $h$ of
the chiral model equation (\ref{chiral}), or alternatively any solution $\chi$
of (\ref{single}), we automatically obtain a solution of the original \nr
self-dual Chern-Simons equations:
\beq
\Psi^{(0)} = \sqrt{{\kappa}}\chi \hskip .75in A_+ ^{(0)}=\chi \hskip .75in
A_-^{(0)}=-\chi ^{\dag}
\label{special}
\eeq
The chiral model equations are also referred to as the ``harmonic map
equations'' because if we regard $J_\pm =h^{-1}\partial_\pm h$ as a connection,
then it satisfies both
\barr
\partial_+ J_- + \partial_- J_+&=&0\cr
\partial_+ J_- - \partial_- J_+ + [ J_+ , J_- ]&=&0
\label{harmonic}
\earr
and so has zero divergence and zero curl.

The {\it global} condition which permits the classification of solutions to the
chiral model equation (\ref{chiral}) is the condition of {\it finiteness} of
the chiral model ``action functional'' (also referred to in the literature as
the ``energy functional'')
\beq
{\cal{E}}[h] \equiv -{1\over 2} \int d^2x~tr(h^{-1}\partial_- h h^{-1}
\partial_+
h)\
\label{chiralenergy}
\eeq
This finiteness condition has direct physical relevance in the related
\nr \cs system because
\barr
{\cal{E}}[h] &=& 2\int d^2 x~tr(\chi\chi^{\dag}) \cr
&=& {2\over\kappa} \int d^2x~tr(\Psi \Psi^{\dag})\cr
&=&{2\over\kappa} {\cal Q}^0
\label{charge}
\earr
where ${\cal Q}^0$ is the conserved gauge invariant matter charge in
(\ref{abcurrent}). Thus, the finite action solutions of the chiral model
equations correspond precisely to the finite charge solutions of the \nr \sd
\cs equations.

In addition to being physically significant, this finiteness condition is
mathematically crucial because it permits the chiral model solutions on ${\bf
R}^2$ to be classified by conformal compactification to the sphere $S^2$
\cite{uhlenbeck,ward2}. Indeed, Uhlenbeck has classified all finite action
chiral model solutions for $SU(N)$ in terms of ``uniton'' factors (which will
be discussed below).

Before discussing the general classification of finite charge solutions, we
introduce the simplest such solutions, the ``single unitons'', upon which the
general solutions are constructed. A ``single uniton'' solution, $h$, of the
$SU(N)$ chiral model equation (\ref{chiral}) has the form
\barr
h=2p-{\bf 1}
\earr
where $p$ is a ``holomorphic projector'' satisfying:
\barr
p^\dag &=&p\\
p^2&=&p\\
({\bf 1} - p ) \partial_+ p & = &0
\label{conditions}
\earr
These single uniton solutions are fundamental to the chiral model system
because as a consequence of the conditions (\ref{conditions}) we find that
\barr
h^{-1}\partial_\pm h &=& \pm 2\partial_\pm p
\earr
{}From this it immediately follows that $h$ satisfies the chiral model equation
(\ref{chiral}). These single uniton solutions are {\it also} solutions of the
$CP^{N-1}$ model since $h$ satisfies the additional $CP^{N-1}$ condition,
$h^2={\bf 1}$, as a result of $p$ being a projector. In terms of the $\chi$
field defined in (\ref{chifield}), the single uniton solutions take the simple
form
\barr
\chi = \partial_+ p
\earr
It is straightforward to check that, as a consequence of the conditions
(\ref{conditions}) satisfied by $p$, $\chi$ satisfies the equation
(\ref{single}), and therefore gives a solution to the \nr \sd \cs equations as
in (\ref{special}).

The general holomorhic projector satisfying the conditions (\ref{conditions})
can be expressed as
\barr
p=M \left ( M^\dag M \right )^{-1} M^\dag
\label{projector}
\earr
where $M$ is any (rectangular) matrix such that $M=M(x^-)$ \cite{uhlenbeck}. It
is easy to see that such an $M$ is a {\it hermitean} projector. The third
condition (\ref{conditions}) is equivalent to $\partial_+ p \,p=0$, which
follows immediately from the fact that
\barr
\partial_+ p = M \left ( M^\dag M \right )^{-1} \partial_+ M^\dag \left ( {\bf
1} - p \right )
\earr

The next step towards the construction of general solutions involves the
process of ``composing'' uniton solutions, as follows. Suppose $h_1=2p_1-{\bf
1}$ is a single uniton solution with $p_1$ satisfying the conditions
(\ref{conditions}) for a holomorphic projector. Further, let $h_2=2p_2-{\bf 1}$
be such that $p_2=p_2^\dag$ and $p_2^2=p_2$. Then $h=h_1 h_2$ is a solution of
the chiral model equation (\ref{chiral}) provided the following first-order
algebro-differential conditions are met:
\barr
& &(i) \hskip .5in \left({\bf 1}-p_2\right)\left(\partial_+ +{1\over 2}
h_1^{-1}\partial_+ h_1 \right) p_2 = 0\cr
& &(ii) \hskip .5in \left({\bf 1}-p_2\right)\left({1\over 2} h_1^{-1}\partial_-
h_1 \right)p_2 = 0
\label{cond}
\earr
Given these conditions,
\barr
h^{-1}\partial_\pm h = \pm 2 \left( \partial_\pm p_1 +\partial_\pm p_2 \right)
\earr
and so, once again, the chiral model equation (\ref{chiral}) is immediately
satisfied.

This procedure of composing uniton-type solutions can be continued, but since
the $p$ matrices involved are projectors, there is a limit to how many
independent projections can be made. For $SU(N)$, at most $N-1$ such terms can
be combined in this manner, as expressed in Uhlenbeck's theorem:
\b

\n{\bf THEOREM} (K.~Uhlenbeck \cite{uhlenbeck}; see also J.~C.~Wood
\cite{wood}): {\it Every finite action solution h of the SU(N) chiral model
equation (\ref{chiral}) may be uniquely factorized as a product of ``uniton''
factors}
\beq
h=\pm h_0 \prod_{i=1}^m (2 p_i -1)\
\label{fact}
\eeq
{\it where:}

\n{\it a) $h_0 \in SU(N)$ is constant;}

\n{\it b) each $p_i$ is a Hermitean projector ($p_i^{\dag}=p_i$ and
$p_i^2=p_i$);}

\m

\n{\it c) defining $h_j=h_0 \prod_{i=1}^j (2p_i-1)$, the following linear
relations must hold:}
\barr
(1-p_i) \left(\partial_+ + {1\over 2} h_{i-1}^{-1} \partial_+
h_{i-1}\right)~p_i&=&0\nonumber\cr
(1-p_i)~h_{i-1}^{-1} \partial_- h_{i-1}~p_i &=&0\cr
\label{linear}
\earr
\n{\it d) $m\leq N-1$.}

\b

\n The $\pm$ sign in (\ref{fact}) has been inserted to allow for the fact that
Uhlenbeck and Wood considered the gauge group $U(N)$ rather than $SU(N)$.

An important implication of this theorem is that for $SU(2)$ {\it all} finite
action solutions of the chiral model have the ``single uniton'' form
\beq
h=(2p-{\bf 1})\
\label{uniton}
\eeq
with $p$ being a holomorphic projector satisfying the conditions
(\ref{conditions}). These single uniton solutions are essentially the $CP^1$
model solutions \cite{din,zak1}. For $SU(N)$ with $N\geq 3$ one must consider
composite unitons in addition to the single unitons. It becomes increasingly
difficult to give a simple characterization of all possible projectors
satisfying the subsidiary linear conditions specified in Uhlenbeck's
construction. However, Wood has presented a systematic parametrization of these
higher unitons, for any $SU(N)$, in terms of a sequence of projectors into
Grassmannian factors. A detailed analysis of the $SU(3)$ and $SU(4)$ cases is
also given in \cite{zak2}.

At this point, it is important to ask how these multi-uniton solutions to the
chiral model equations relate to the special explicit Toda-type solutions
discussed previously in (\ref{toda}-\ref{suntoda}). While the algebraic {\it
Ans\"atze} (\ref{ans1},\ref{ans2}) each leads to a non-Abelian charge density
$\rho=[\Psi^{\dag}, \Psi]$ which is {\it diagonal}, the chiral model solutions
(\ref{special}) have charge density $\rho^{(0)} = {\kappa}[\chi^{\dag},
\chi]$ which need not be diagonal. However, $\rho$ is always hermitean, and so
it can be diagonalized by a gauge transformation. It is still a nontrivial
algebraic task to implement this diagonalization explicitly, but this can be
done for the $SU(N)$ solutions, revealing an interesting new link between the
chiral model and the Toda system \cite{dunne3}.

It is instructive to illustrate this procedure with the $SU(2)$ case first.
Here, Uhlenbeck's theorem implies that the only finite charge solution has the
form $\chi=\partial_+ p$, where $p$ is a holomorphic projector as in
(\ref{projector}). For $SU(2)$ we can only project onto a {\it line} in ${\bf
C}^2$, so we take
\beq
M(x^-)=\left(\matrix{1\cr f(x^-)\cr}\right)\
\label{mvec}
\eeq
This then leads to the projection matrix
\beq
p={1\over 1+f \bar{f}} \left(\matrix{1&\bar{f}\cr f& f \bar{f}\cr} \right)
\eeq
and the corresponding solution $\chi$ can be expressed in terms of the single
holomorphic function $f(x^-)$ :
\beq
\chi = \partial_+p = {f\partial_+ \bar{f} \over (1+f\bar{f})^2}
\left(\matrix{-1&1/f \cr-f&1\cr}\right)\
\label{chi}
\eeq
The corresponding matter density is
\beq
[\chi^{\dag}, \chi] = -{\partial_+ \bar{f} \partial_- f\over (1+f\bar{f})^3}
\left(\matrix{1-f\bar{f}&2\bar{f} \cr 2f&-1+f\bar{f} \cr}\right)\
\label{dens}
\eeq
which may be diagonalized by the unitary matrix
\barr
g&=&{1\over \sqrt{1+f\bar{f}}} \left(\matrix{-\bar{f}&1\cr
1&f\cr}\right)
\earr
to yield the diagonalized charge density
\barr
g^{-1}[\chi^{\dag}, \chi] g&=&\partial_+ \partial_- {\rm
log}(1+f(x^-)\bar{f}(x^+))\left(\matrix{1&0\cr 0&-1\cr}\right)\cr
&=&\partial_+ \partial_- {\rm log~det}(M^{\dag}
M)~\left(\matrix{1&0\cr 0&-1\cr}\right)
\label{diag}
\earr
In this diagonalized form we recognize Liouville's solution (\ref{liou-sol}) to
the classical $SU(2)$ Toda equation. It is worth emphasizing that for $SU(2)$
the \nr \sd \cs equations (\ref{nrsd1},\ref{nrsd2}) can be converted, by
suitable algebraic ansatze as discussed in the previous section, into the
classical Toda (i.e. Liouville) equation or the affine Toda (i.e. sinh-Gordon)
equation. However, the above analysis shows that only the classical Toda case
(i.e. Liouville) corresponds to finite charge.

A similar construction is possible for the $SU(N)$ case \cite{dunne3,dunne4}.
Specifically, let
\beq
h=(-1)^{{1\over 2}N(N+1)} \prod_{a=1}^{N-1} (2p_{a} -{\bf 1})\
\label{sunh}
\eeq
be a product where each $p_a$ is a holomorphic projector onto the
$a$-dimensional subspace spanned by the columns of the $N\times a$ rectangular
matrix $M_a(x^-)$ in (\ref{suntoda},\ref{uvec}):
\barr
M_a=\left(\matrix{1&0&\dots&0\cr
f_1&\partial_-f_1&\dots&\partial_-^{(a-1)}f_1\cr
f_2&\partial_-f_2&\dots&\partial_-^{(a-1)}f_2\cr
\vdots&\vdots& &\vdots\cr
f_{N-1}&\partial_-f_{N-1}&\dots&\partial_-^{(a-1)}f_{N-1}}\right)
\label{mmatrix}
\earr
Then $h$ is a finite action solution of the $SU(N)$ chiral model equation
(\ref{chiral}) and the corresponding solution of the \nr \sd \cs equations is
\barr
\chi = \sum_{a=1}^{N-1} \partial_+ p_a
\earr
The charge density $[\chi^{\dag}, \chi]$ may be diagonalized by an $SU(N)$
gauge transformation $g$ yielding a diagonal form
\beq
g^{-1}[\chi^{\dag}, \chi] g=\sum_{a =1}^{N-1} \{\partial_+ \partial_- {\rm
log~det}(M_{a}^{\dag} M_{a})\} H_{a}\
\label{sundiag}
\eeq
where the $H_{a}$ are the Cartan subalgebra generators of $SU(N)$ in the
Chevalley basis. This diagonal form of the charge density corresponds precisely
to the $SU(N)$ Toda solution (\ref{suntoda}).

Another useful result which follows from the relationship between the \nr \sd
\cs equations (\ref{nrsd1},\ref{nrsd2}) and the chiral model equation
(\ref{chiral}) is that the chiral model energy (\ref{chiralenergy}) is
quantized in integral multiples of $8\pi$ \cite{valli}. This implies that the
abelian Chern-Simons charge ${\cal Q}^0\equiv\int tr(\Psi^{\dag} \Psi)$ is
quantized in integral multiples of $4\pi \kappa$. A related quantization
condition has been noted in \cite{dunne1}, where the {\it non-Abelian} charges
${\cal Q}^0_a\equiv\int\rho_a$ are quantized in integral multiples of $4\pi
\kappa$ for the $SU(N)$ Toda-type solutions (\ref{suntoda}). In this case the
abelian charge is the sum of the individual nonAbelian charges : ${\cal
Q}^0=\sum_{a}{\cal Q}^0_a$.

\section{Relativistic SDCS Theories}

\b

\n{3.1 {\it Relativistic Self-Dual Chern-Simons Equations}}

\b

In this section we discuss the relativistic generalization of the \nr \sd \cs
theories. The existence of vortex solutions in $2+1$-dimensional relativistic
gauge-Higgs models including Chern-Simons terms has been known for some time
\cite{khare}. The importance of self-duality was first noticed in the context
of {\it abelian} theroies \cite{hong,jackiw3}, where vortices in the
relativistic Chern-Simons-Higgs model were shown to be related to a
self-duality condition reminiscent of Bogomol'nyi's analysis \cite{bog} of
vortices in the abelian Higgs model. With a particular sixth order scalar
potential there is a lower bound on the energy functional which is saturated by
topological solitons and nontopological vortices \cite{jackiw4}. An extension
of these abelian models is possible, to nonabelian \r \sd \cs theories with a
global $U(1)$ symmetry \cite{klee1}, once again with a special sixth order
potential. However, while the self-dual structure of the model generalizes in a
relatively straighforward manner, the analysis of the nonabelian \r \sd \cs
equations themselves is significantly more complicated, and correspondingly
more interesting. The richness of the nonabelian theory is compounded by the
many available choices: of gauge group, of representation, of matter coupling,
etc... \cite{klee1,klee2,dunne5,gustavo}. Matter fields in the defining
representation have been studied in \cite{klee2}, while the most interesting
case once again seems to be the case of adjoint coupling
\cite{klee1,dunne5,kao1,dunne6}. The self-dual structure of these \r \sd
Chern-Simons systems is related at a fundamental level to extended
supersymmetry in $2+1$ dimensions \cite{clee,ivanov,kao2}, in the sense that
the self-dual Lagrangian is the bosonic portion of a Lagrangian with an
extended supersymmetry. This is in accordance with a general relationship
between self-duality and extended supersymmetry \cite{hlousek}.

Consider the Lagrange density
\beq
{\cal L}=-\kappa {\cal L}_{CS}-tr \left(\left(D_\mu \phi\right)^\dag D^\mu
\phi\right) - V\left(\phi, \phi^\dag\right)
\label{rlag}
\eeq
where ${\cal L}_{CS}$ is the \cs Lagrange density in (\ref{cslag}), and the
scalar field potential $V(\phi, \phi^\dag)$ is
\beq
V\left(\phi, \phi^\dag\right) = {1\over 4\kappa^2}tr\left( \left([\;[\;\phi,
\phi^\dag\;],\phi\;]-v^2\phi\right)^\dag\;\left([\;[\;\phi,
\phi^\dag\;],\phi\;]-v^2\phi\right)\right).
\label{rpot}
\eeq
The space-time metric is taken to be $g_{\mu\nu}={\rm diag}\left( -1,1,1
\right)$ and, as before, $tr$ refers to the trace in a finite dimensional
representation of the compact simple Lie algebra ${\cal G}$ to which the gauge
fields $A_\mu$ and the charged matter fields $\phi$ and $\phi^\dag$ belong. The
$v^2$ parameter appearing in the potential (\ref{rpot}) will play the role of a
mass parameter (see (\ref{mass})). Under a gauge transformation both the
potential $V$ and the scalar field kinetic term  $tr\left(\left(D_\mu
\phi\right)^\dag D^\mu \phi\right)$ remain invariant. However, the Chern-Simons
Lagrange density is not invariant and the dimensionless coupling coefficient
$\kappa$ must be quantized in order for the corresponding quantum theory to be
invariant under large gauge transformations \cite{deser1}. The particular {\it
sixth}-order form of the scalar field potential (\ref{rpot}), together with its
overall strength depending on the \cs coupling parameter $\kappa$, are fixed by
the condition of self-duality, as shown below.

The Euler-Lagrange equations of motion obtained from the Lagrange density
(\ref{rlag}) are:
\barr
D_\mu D^\mu \phi&=&{\partial V \over \partial \phi^\dag}
\label{rmatter}
\earr
\barr
- \kappa \epsilon^{\mu \nu \rho}F_{\nu \rho}& = &i J^{\mu}
\label{rgauge}
\earr
In the matter equation of motion (\ref{rmatter}), ${\partial V \over \partial
\phi^\dag}$ is defined by the change in the potential $V$ under a variation of
$\phi^\dag$:
\beq
\delta V \equiv tr\left(\delta\phi^\dag{\partial V \over \partial
\phi^\dag}\right)
\label{variation}
\eeq
In the gauge equation of motion (\ref{rgauge}), $J^\mu$ is the relativistic
nonabelian current
\beq
J^\mu \equiv -i\left([\;\phi^\dag, D^\mu \phi\;] - [\;(D^\mu\phi)^\dag,
\phi\;]\right)
\label{rcurrent}
\eeq
which is covariantly conserved : $D_\mu J^\mu =0$. This system also has an
abelian current, $Q_\mu$,
\beq
Q_\mu = -i \, tr\left(\phi^\dag D_\mu \phi - \left( D_\mu \phi\right)^\dag
\phi\right),
\label{rabeliancurrent}
\eeq
which is ordinarily conserved : $\partial_\mu Q^\mu =0$.

The energy density corresponding to the Lagrange density (\ref{rlag}) is
\barr
{\cal H}=tr\left(\left(D_0\phi\right)^\dag D_0\phi\right)+tr\left(\left(
D_i\phi\right)^\dag D_i\phi\right)+V\left(\phi,\phi^\dag\right),
\label{rham}
\earr
supplemented by the Gauss law constraint
\beq
[\phi^\dag , D_0\phi ] - [ \left( D_0\phi \right)^\dag , \phi ] = 2 \kappa F_{1
2},
\label{rgauss}
\eeq
which is the zeroth component of the gauge field equations of motion
(\ref{rgauge}). Notice that, as is familiar for Chern-Simons theories, the
Chern-Simons term ${\cal L}_{CS}$ in the Lagrange density (\ref{rlag}) does not
contribute to the energy, while it does affect the canonical structure and the
constraints \cite{deser1,witten,dunne1}.

To find self-dual solutions which minimize the energy, we re-express the energy
density in a modified form, using an adaptation of the Bogomol'nyi method for
vortices in the abelian Higgs model \cite{bog}. Using the identity
(\ref{bochner}) together with the Gauss law constraint (\ref{rgauss}), we can
write
\barr
tr\left(\left( D_i\phi\right)^\dag D_i\phi\right)&=&tr\left(\left(
D_-\phi\right)^\dag D_-\phi\right)\cr\cr
&& +{i\over 2\kappa} tr\left( \left( [\;[\;\phi,\phi^\dag\;],\phi
\;]\right)^\dag D_0\phi - [\;[\;\phi,\phi^\dag\;],\phi\;] \left(
D_0\phi\right)^\dag \right)\cr
&&
\label{bogol2}
\earr
where we recall that $D_\pm \equiv D_1\pm i D_2$. The second term on the RHS of
(\ref{bogol2}) may be cancelled in the energy density (\ref{rham}) by a term
from $tr\left(\left( D_0\phi\right)^\dag D_0\phi\right)$ if we write
\barr
tr\left(\left( D_0\phi\right)^\dag D_0\phi\right) &=&tr\left(\left(
D_0\phi-{i\over 2\kappa}[\,[\,\phi,\phi^\dag\,],\phi\,]\right)^\dag\left(
D_0\phi-{i\over 2\kappa}[\,[\,\phi,\phi^\dag\,],\phi\,]\right)\right) \cr\cr
& &- {i\over 2\kappa} tr\left( \left( [\,[\,\phi,\phi^\dag\,],\phi
\,]\right)^\dag D_0\phi - [\,[\,\phi,\phi^\dag\,],\phi\,] \left(
D_0\phi\right)^\dag \right)\cr\cr
& &- {1\over 4\kappa^2}tr\left( \left([\,[\,\phi,
\phi^\dag\,],\phi\,]\right)^\dag\,[\,[\,\phi, \phi^\dag\,],\phi\,]\right)
\label{dzero1}
\earr
One could then interpret the final term on the RHS of (\ref{dzero1}) as (minus)
a  potential, in which case the energy density (\ref{rham}) could be expressed
in a manifestly positive form. However, this choice would result in a sixth
order scalar field potential {\it without} a mass term, and this is unsuitable
for a number of reasons discussed below. Rather, one should be more general and
explicitly introduce a mass term in this potential by writing (generalizing the
decomposition (\ref{dzero1}))
\barr
& &tr\left(\left( D_0\phi\right)^\dag D_0\phi\right)=\cr\cr
& &~~~tr\left(\left( D_0\phi-{i\over
2\kappa}\left([\,[\,\phi,\phi^\dag\,],\phi\,]-v^2 \phi\right)\right)^\dag\left(
D_0\phi-{i\over 2\kappa}\left([\,[\,\phi,\phi^\dag\,],\phi\,]-v^2
\phi\right)\right)\right) \cr\cr
& &~~~- {i\over 2\kappa} tr\left(\left( [\,[\,\phi,\phi^\dag\,],\phi\,]-v^2
\phi\right)^\dag D_0\phi - \left( [\,[\,\phi,\phi^\dag\,],\phi\,] -v^2
\phi\right)\left( D_0\phi\right)^\dag\right) \cr\cr
& &~~~- {1\over 4\kappa^2}tr\left( \left([\,[\,\phi,
\phi^\dag\,],\phi\,]-v^2\phi\right)^\dag\,\left([\,[\,\phi,
\phi^\dag\,],\phi\,]-v^2\phi\right)\right)
\label{dzero2}
\earr
The final term in this expression (\ref{dzero2}) is recognized as (minus) the
potential $V\left(\phi , \phi^\dag\right)$ defined in (\ref{rpot}), and so the
energy density (\ref{rham}) can be expressed as
\barr
{\cal E} &=& tr\left(\left( D_0\phi-{i\over
2\kappa}\left([\,[\,\phi,\phi^\dag\,],\phi\,]-v^2 \phi\right)\right)^\dag\left(
D_0\phi-{i\over 2\kappa}\left([\,[\,\phi,\phi^\dag\,],\phi\,]-v^2
\phi\right)\right)\right)\cr\cr
& & + tr\left(\left( D_-\phi\right)^\dag D_-\phi\right) +{iv^2 \over 2\kappa
}tr \left( \phi^\dag \left( D_0\phi\right) -\left( D_0 \phi\right)^\dag \phi
\right)
\label{factor}
\earr
The first two terms in (\ref{factor}) are manifestly positive and the third
gives a lower bound for the energy density, which may be written in terms of
the time component, $Q^0$, of the abelian relativistic current defined in
(\ref{rabeliancurrent}):
\beq
{\cal E} \geq {v^2 \over 2\kappa} Q^0
\label{rbound}
\eeq
This lower bound (\ref{rbound}) is saturated when the following two conditions
(each first order in spacetime derivatives) hold:
\barr
D_- \phi&=&0
\label{condition1}
\earr
\barr
D_0 \phi&=&{i\over 2\kappa} \left( [\;[\;\phi,\phi^\dag\;],\phi
]-v^2\phi\right)
\label{condition2}
\earr
The consistency condition of these two equations states that
\barr
\left( D_0 D_- -D_- D_0 \right) \phi &\equiv& [F_{0-},\phi ]\cr
&=&-{i\over 2\kappa} [[ \phi , (D_+ \phi )^\dag ], \phi ]\cr
&=&{1\over 2 \kappa} [ J_- , \phi ]
\label{consistency}
\earr
which expresses the gauge field Euler-Lagrange equation of motion
\barr
F_{0-}&=&{1\over 2\kappa}J_-\cr
&=&-{1\over 2\kappa}[ \phi , \left(D_+\phi\right)^\dag ]
\label{spatial}
\earr
for the spatial components of the current. The other gauge field equation,
$F_{+-}={1\over \kappa} J_0$, may be re-expressed using equation
(\ref{condition2}) in a form not involving explicit time derivatives. We thus
arrive at the {\it relativistic self-dual Chern-Simons
equations}:
\barr
D_- \phi &=&0
\label{rsd1}
\earr
\barr
F_{+-} &=& {1\over \kappa^2} [ v^2 \phi -[  [ \phi, \phi^\dag ], \phi]
,\phi^\dag  ]
\label{rsd2}
\earr
At the self-dual point, we can use equation (\ref{condition2}) to express the
energy density as
\beq
{\cal E}_{\rm SD} = {v^2\over 2\kappa^2} tr \left(\phi^\dag \left (v^2 \phi-[
[ \phi,
\phi^\dag ], \phi] \right)\right)
\label{sdbound}
\eeq
Recall that all solutions to the {\it nonrelativistic} self-duality equations
(\ref{nrsd1},\ref{nrsd2}) correspond to the static zero-energy solutions to the
Euler-Lagrange equations of motion \cite{dunne1}. Here, in the relativistic
theory, the situation is rather different. First, the lower bound
(\ref{rbound}) on the energy density is not necessarily zero, and the solutions
of (\ref{condition2}) are time dependent. Furthermore, unlike in the
nonrelativistic case, it is possible to have nontrivial solutions for $\phi$
while having $F_{+-}=0$. These solutions do have zero energy, and are gauge
equivalent to solutions of the {\it algebraic} equation
\beq
[ [ \phi, \phi^\dag ], \phi ] =v^2 \phi .
\label{min}
\eeq
Solutions of this equation also correspond to the minima of the potential
(\ref{rpot}), and these potential minima are clearly degenerate.

A class of solutions to the self-duality equations (\ref{rsd2}) is given by the
following zero energy solutions of the Euler-Lagrange equations:
\barr
\phi&=&g^{-1} \phi_{(0)} g  \cr
A_\pm &=& g^{-1} \partial_\pm g \cr
A_0&=&g^{-1} \partial_0 g
\label{zeroenergy}
\earr
where $\phi_{(0)}$ is any solution of (\ref{min}), and $g=g(\vec{x},t)$ takes
values in the gauge group. It is clear that these solutions satisfy $D_0
\phi=0$, $D_-\phi=0$, $F_{+-}=0$, as well as the algebraic equation
(\ref{min}), which implies that they are self-dual, and that they have zero
magnetic field and zero charge density. While this class of solutions may look
somewhat trivial, it is still important because the solutions, $\phi_{(0)}$, of
the algebraic equation (\ref{min}) classify the minima of the potential $V$,
and the finite {\it nonzero} energy solutions of the self-duality equations
must be gauge equivalent to such a solution at infinity:
\beq
\phi \to g^{-1} \phi_{(0)} g \quad\quad {\rm as} \quad r\to\infty
\eeq

It is important to check explicitly the consistency of the self-duality
equations (\ref{rsd1},\ref{rsd2}) with the Euler-Lagrange equation of motion
(\ref{rmatter},\ref{rgauge}). Note that
\barr
D_\mu D^\mu \phi &=& -D_0 D_0 \phi + D_i D_i \phi \cr\cr
&=& -D_0 D_0 \phi + D_+ D_- \phi + i [F_{12}, \phi]
\label{deldel}
\earr
For self-dual solutions $D_- \phi=0$, and using the self-duality equation for
$D_0 \phi$ we find that
\barr
i [F_{12},\phi]={1\over 2 \kappa^2 }[[\phi,[\phi^\dag,[\phi,\phi^\dag]]],\phi
]+{v^2\over 2 \kappa^2 }[\phi,[\phi,\phi^\dag ]]
\label{phicheck1}
\earr
and
\barr
-D_0 D_0 \phi = {v^4\over 4 \kappa^2} \phi +{v^2\over
2\kappa^2}[\phi,[\phi,\phi^\dag ]] +{1\over 4 \kappa^2}
[[\phi,[\phi,\phi^\dag]],[\phi,\phi^\dag]]
\label{phicheck2}
\earr
Therefore,
\barr
D_\mu D^\mu \phi &=& {v^4\over 4 \kappa^2} \phi +{v^2\over
\kappa^2}[\phi,[\phi,\phi^\dag ]] \cr\cr
& &+{1\over 4 \kappa^2}\left( [[\phi,[\phi,\phi^\dag]],[\phi,\phi^\dag]] +2
[[\phi,[\phi^\dag,[\phi,\phi^\dag]]],\phi ]\right)
\label{phicheck3}
\earr
It is a straightforward matter to verify that (\ref{phicheck3}) does indeed
yield the correct charged scalar field Euler-Lagrange equation of motion
(\ref{rmatter}) with the potential $V\left(\phi, \phi^\dag \right)$ given by
equation (\ref{rpot}).

To verify that this model is the natural nonabelian generalization of the
abelian relativistic model \cite{hong,jackiw3}, we take the `abelian limit' by
choosing a special algebraic restriction of $SU(2)$. (Such an abelian limit is
familiar from the corresponding nonrelativistic models \cite{dunne2}). Consider
the Chevalley basis for the $SU(2)$ Lie algebra generators:
\barr
[E_+,E_-]&=& H\cr
[H,E_\pm ]&=&\pm 2 E_\pm \cr
tr\left( E_+ E_- \right) &=&1 \cr
tr\left( H^2 \right) &=&2
\label{su2chevalley}
\earr
where $H$ is the Cartan subalgebra generator and $E_\pm$ are the step
operators. For example, in the defining representation of $SU(2)$, this basis
may be taken as:
\barr
E_+=\left(\matrix{0&1\cr 0&0}\right) \hskip .5in E_-=\left(\matrix{0&0\cr
1&0}\right) \hskip .5in H=\left(\matrix{1&0\cr 0&-1}\right)
\earr
Further, choose the fields to have the following Lie algebraic decomposition
(note that this is an {\it ansatz}, not simply a gauge choice):
\barr
\phi &=& \psi E_+\cr
\phi^\dag &=& \bar{\psi} E_- \cr
A_-&=&a H\cr
A_+&=&-\bar{a} H
\label{example}
\earr
Then $D_- \phi =\left(\partial_- \psi +2 a\psi\right) E_+$, and the
self-duality equations (\ref{rsd1},\ref{rsd2}) become
\barr
a&=&-{1\over 2}\partial_-{\rm ln} \psi\cr
\partial_+ a + \partial_- \bar{a}&=&-{1\over \kappa^2 } |\psi |^2\left( 2|\psi
|^2-v^2\right)
\label{exdual}
\earr
These two equations may be combined to yield the single equation satisfied by
the gauge invariant scalar, $|\psi |^2 \equiv tr\left( \phi \phi^\dag\right) $:
\beq
\partial_+ \partial_- {\rm ln} |\psi |^2 = {2\over \kappa^2} |\psi |^2\left(
2|\psi |^2-v^2\right)
\label{abelian}
\eeq
This is the same (apart from trivial rescalings resulting from different
normalization) as the abelian self-duality condition found in
\cite{hong,jackiw3} in their analysis of abelian self-dual Chern-Simons
vortices. With the Lie algebraic {\it ansatz} (\ref{example}) for the fields,
the potential (\ref{rpot}) reduces to
\barr
V = {1\over 4\kappa^2} |\psi |^2 \left(2|\psi |^2 -v^2\right)^2
\earr
which is the same as the self-dual sixth-order potential found in
\cite{hong,jackiw3}. Further, with the fields as in (\ref{example}), the
self-dual  energy density (\ref{sdbound}) becomes
\barr
{\cal E}_{\rm SD}&=&-{v^2\over  2\kappa^2} |\psi |^2 \left( 2|\psi
|^2-v^2\right)
\earr
{}From (\ref{exdual}) we recognize this self-dual energy density as being
proportional to the `abelian' magnetic field strength, $f_{+-}\equiv{1\over
2}tr\left( H F_{+-}\right)=\partial_+ a +\partial_- \bar{a}$, and so
\barr
{\cal E}_{\rm SD} = {v^2\over 2 } f_{+-}
\earr
which shows that the energy is bounded below by a magnetic flux, as in the
abelian model \cite{hong,jackiw3}.

We note here that the positive sign of the RHS of (\ref{abelian}) is
significant. For example, in the {\it massless} case when $v^2=0$, the equation
\beq
\partial_+ \partial_- {\rm ln} |\psi |^2 = {4\over \kappa^2} |\psi |^4
\label{case}
\eeq
can be solved exactly, but it has no real, regular and integrable solutions for
$|\psi |^2$. This lack of real, regular solutions when $v^2=0$ is one reason
for introducing the $v^2$ mass term in the potential (\ref{rpot}).

Another reason for introducing the mass term in the potential is that it
permits the taking of the nonrelativistic limit. Without a mass scale for the
scalar field $\phi$ there is no meaning to such a limit. Restoring factors of
$c$ to the potential (\ref{rpot}), we find a mass term \cite{dunne5}
\beq
{v^4\over 4\kappa^2 c^4}tr\left(\phi^\dag \phi\right) \equiv m^2 c^2
tr\left(\phi^\dag \phi\right)
\label{massterm}
\eeq
which yields a scalar field mass
\beq
m={v^2\over 2 \kappa c^3}
\label{mass}
\eeq
To maintain a finite mass in the nonrelativistic limit, the $c\to\infty$ limit
must be accompanied by the $v^2\to\infty$ limit in such a way that ${v^2\over
c^3}$ is kept constant. Separating out the rest-mass energy as
\beq
\phi = {1\over \sqrt{2m}} e^{-imc^2 t}\Psi ,
\label{nonreldecomp}
\eeq
and keeping only the dominant terms in inverse powers of $c$, the Lagrange
density (\ref{rlag}) reduces to the nonrelativistic Lagrange density (the
Chern-Simons Lagrange density, ${\cal L}_{\rm CS}$, is unchanged) in
(\ref{nrlag}) :
\beq
{\cal L}=-\kappa {\cal L}_{\rm CS}+i\, tr\left(\Psi^\dag \left(\partial_t
\Psi+[A_0, \Psi ]\right)\right)-{1\over 2m}tr\left(\left( D_i\Psi\right)^\dag
D_i\Psi\right) +{1\over 4m\kappa c} tr\left(\left( [\Psi,\Psi^\dag]\right)
^2\right)
\label{nonrellag}
\eeq
Further, in the nonrelativistic limit, the relativistic self-duality equations
(\ref{rsd1},\ref{rsd2}) reduce to
\barr
D_- \Psi &=&0\cr
F_{+-}&=&{1\over \kappa } [\Psi,\Psi^\dag ]
\label{nrsd}
\earr
which are the nonrelativistic self-dual Chern-Simons equations
(\ref{nrsd1},\ref{nrsd2}). Finally, at the self-dual point (where $D_-\Psi=0$),
the Schr\"odinger equation (\ref{matter}) becomes
\beq
i D_t \Psi = -{1\over 4m\kappa c} [[\Psi,\Psi^\dag],\Psi]
\eeq
which is the nonrelativistic limit of the relativistic self-duality equation
(\ref{condition2}).

\b

\n{3.2 {\it Classification of Minima}}

\b

The sixth order potential (\ref{rpot}) has degenerate minima given by fields
$\phi_{(0)}$ which solve
\barr
[\,[\,\phi, \phi^\dag\,], \phi\,]=\phi
\label{minmin}
\earr
where a factor of $v$ has been absorbed into the field $\phi$. We recognize the
condition (\ref{minmin}) as the $SU(2)$ commutation relation. For a
general gauge algebra, finding the solutions to (\ref{minmin}) is the classic
Dynkin problem \cite{dynkin} of embedding $SU(2)$ into a general Lie
algebra. It is interesing to note that this type of embedding problem
also plays a significant role in the theory of spherically symmetric magnetic
monopoles and the Toda molecule equations \cite{leznov2}.

It is clear that in order to satisfy (\ref{minmin}) for a general gauge
algebra, $\phi=\phi_{(0)}$ must be a linear combination of the step operators
for the {\it positive} roots of the algebra. Further, since we have the freedom
of global gauge invariance, we can choose representative gauge inequivalent
solutions $\phi_{(0)}$ to be linear combinations of the step operators of the
positive {\it simple} roots. It is therefore convenient to work in the
Chevalley basis (\ref{chevalley}) for the gauge algebra (for ease of
presentation we shall consider $SU(N)$). Expand $\phi_{(0)}$ in terms of the
positive simple root step operators as:
\beq
\phi_{(0)} = \sum_{a=1}^{N-1} \phi_{(0)}^a E_a
\label{expansion}
\eeq
Then $[\phi_{(0)}, \phi_{(0)}^\dag ]$ is diagonal,
\beq
[\phi_{(0)}, \phi_{(0)}^\dag ] = \sum_{a=1}^{N-1} |\phi_{(0)}^a |^2 H_a .
\eeq
The Chvalley basis commutation relations (\ref{chevalley}) then imply that
\barr
[[\phi_{(0)}, \phi_{(0)}^\dag ], \phi_{(0)}] = \sum_{a=1}^{N-1}
\sum_{b=1}^{N-1} |\phi_{(0)}^a |^2 \phi_{(0)}^b C_{b\, a}E_b
\earr
which, like $\phi_{(0)}$, is once again a linear combination of just the simple
root step operators. Thus, for suitable choices of the coefficients
$\phi_{(0)}^a$, it is {\it possible} for the $SU(N)$ algebra element
$\phi_{(0)}$ to satisfy the $SU(2)$ commutation relation $[ [ \phi, \phi^\dag
], \phi ]=\phi$.

For example, one can always choose $\phi_{(0)}$ proportional to a {\it
single} step operator, which by global gauge invariance can always be taken to
be $E_1$ :
\barr
\phi_{(0)}={1\over \sqrt{2}} E_1
\label{first}
\earr
In the other extreme, the $SU(N)$ ``maximal embedding'' case, with {\it all}
$N-1$ step operators involved in the expansion (\ref{expansion}), the solution
for $\phi_{(0)}$ is :\footnote{In general, the squares of the coefficients for
the maximal embedding case are the coefficients, in the simple root basis, of
(one half times) the sum of {\it all} positive roots of the algebra.}
\beq
\phi_{(0)}={1\over \sqrt{2}}\sum_{a=1}^{N-1} \sqrt{a(N-a)}\; E_a
\label{max}
\eeq
All other solutions for $\phi_{(0)}$, intermediate between the two extremes
(\ref{first}) and (\ref{max}), can be generated by the following systematic
procedure. If one of the simple root step operators, say $E_b$, is omitted from
the summation in (\ref{expansion}) then this effectively decouples the $E_{\pm
a}$'s with $a<b$ from those with $a>b$. Then the coefficients for the $(b-1)$
step operators $E_a$ with $a<b$ are just those for the maximal embedding
(see equation (\ref{max})) in $SU(b)$, and the coefficients for the $(N-b-1)$
$E_a$'s with $a>b$ are those for the maximal embedding in $SU(N-b)$:
\barr
\phi_{(0)}={1\over \sqrt{2}}\sum_{a=1}^{b-1} \sqrt{a(b-a)} E_a +{1\over
\sqrt{2}}\sum_{a=b+1}^{N-1} \sqrt{a(N-b-a)} E_a
\label{oneomission}
\earr
Diagrammatically, we can represent the maximal embedding case
(\ref{max}) with the Dynkin diagram of $SU(N)$ :
\barr
\underbrace{o-o-o- \dots -o-o}_{N-1}
\earr
which shows the $N-1$ simple roots of the algebra, each connected to its
nearest neighbours by a single line. Omitting the $b^{th}$ simple root step
operator from the sum in (\ref{expansion}) can be conveniently represented
as breaking the Dynkin diagram in two by deleting the $b^{th}$ dot:
\barr
\underbrace{o-o- \dots -o}_{b-1} - \times - \underbrace{o- \dots
-o}_{N-b-1}
\earr
With this deletion of the $b^{th}$ dot, the $SU(N)$ Dynkin diagram breaks
into the Dynkin diagram for $SU(b)$ and that for $SU(N-b)$. Since the
remaining simple root step operators decouple into a Chevalley basis for
$SU(b)$ and another for $SU(N- b)$, the coefficients required for the
summation over the first $b-1$ step  operators are just those given in
(\ref{max}) for the maximal embedding in $SU(b)$, while the coefficients for
the summation over the last $N-b-1$ step operators are given by the
maximal embedding for $SU(N-b)$, as indicated in (\ref{oneomission}).

It is clear that this process may be repeated with further roots being
deleted from the Dynkin diagram, thereby subdividing the original $SU(N)$
Dynkin diagram, with  its $N-1$ consecutively linked dots, into subdiagrams of
$\leq N-1$ consecutively linked dots. The final diagram, with $M$ deletions
made, can be characterized, up to gauge equivalence, by the $M+1$ lengths of
the remaining consecutive strings of dots. A simple counting argument shows
that the total number of ways of doing this (including the case where {\it all}
dots are deleted, which corresponds to the trivial solution $\phi_{(0)}=0$) is
given by the number, $p(N)$, of (unrestricted) partitions of $N$.

The $SU(4)$ case is sufficient to illustrate this procedure. There are 5
partitions of 4, and they correspond to the following solutions for
$\phi_{(0)}$:
\barr
o-o-o {\hskip 1in}&& \phi_{(0)}={1\over \sqrt{2}}\left( \sqrt{3}E_1
+2E_2+\sqrt{3}E_3
\right)\cr
o-o-\times {\hskip 1in}&& \phi_{(0)}=E_1+E_2\cr
o-\times -o {\hskip 1in}&& \phi_{(0)}={1\over \sqrt{2}} E_1
+{1\over\sqrt{2}} E_3 \cr
o-\times -\times {\hskip 1in}&&\phi_{(0)}={1\over \sqrt{2}}E_1\cr
\times -\times -\times {\hskip 1in}&& \phi_{(0)}=0
\label{su4example}
\earr

Thus we have a simple constructive procedure, and a correspondingly simple
labelling notation, for finding all $p(N)$ gauge inequivalent solutions
$\phi_{(0)}$ to the algebraic embedding condition (\ref{minmin}). Recall that
each
such $\phi_{(0)}$ characterizes a distinct minimum of the potential $V$, as
well as a class of zero energy solutions to the selfduality equations
(\ref{rsd1},\ref{rsd2}).

Since each vacuum solution $\phi_{(0)}$ corresponds to an embedding of
$SU(2)$ into $SU(N)$, an alternative shorthand for labelling the different
vacua consists of listing the block diagonal spin content of the $SU(2)$ Cartan
subagebra element $[\phi_{(0)}, \phi_{(0)}^\dag ]\sim J_3$. For example,
consider the matter fields $\phi$ taking values in the $N\times N$ defining
representation. Then, for each vacuum solution, $[\phi_{(0)},\phi_{(0)}^\dag ]$
takes the $N\times N$ diagonal sub-blocked form:

\barr
[\phi_{(0)}, \phi_{(0)}^\dag ]=\left( \matrix{j_1\cr &\ddots \cr &&-j_1\cr
&&&j_2\cr &&&&\ddots\cr &&&&&-j_2\cr &&&&&&\ddots\cr
&&&&&&&j_M\cr
&&&&&&&&\ddots\cr &&&&&&&&&-j_M}\right)
\earr
Each spin $j$ sub-block has dimension $2j+1$, and so it is therefore natural
to associate this particular $\phi_{(0)}$ with the following partition of $N$ :
\barr
N=(2j_1+1)+(2j_2+1)+\dots +(2j_M+1)
\earr
For example, the $SU(4)$ solutions listed in (\ref{su4example}) may be
labelled by the partitions $4$, $3+1$, $2+2$, $2+1+1$, and $1+1+1+1$,
respectively.

\b

\n{3.3 {\it Vacuum Mass Spectra}}

\b

Having classified all possible gauge inequivalent vacua of the potential
$V$, we now determine the spectrum of massive excitations in each vacuum. In
the abelian model \cite{hong,jackiw3} there is only one nontrivial vacuum, and
a consequence of the particular sixth order self-dual form of the potential is
that in this broken vacuum the massive gauge excitation and the remaining real
massive scalar field are degenerate in mass. This degeneracy of the gauge and
scalar masses in the broken vacuum is also true of the $2+1$ dimensional
Abelian Higgs model \cite{bog}. In the nonabelian models considered here the
situation is considerably more complicated, due to the presence of many fields
and also due to the many different gauge inequivalent vacua. Nevertheless, we
shall see that an analogous mass degeneracy pattern exists, reflecting the
self-dual character of the potential (\ref{rpot}).

Regarded as a symmetry breaking problem, the \r \sd \cs system with Lagrange
density (\ref{rlag}) is rather different from a conventional Higgs system.
First, in $3+1$ dimensional field theory one most commonly considers symmetry
breaking potentials of $\phi^4$ form, but here in $2+1$ dimensions we consider
a (renormalizable) sixth order potential (\ref{rpot}). This means that the
extraction of the scalar masses in the broken vacua is algebraically more
complicated. The second, and more significant, difference is that the Higgs
mechanism for generating massive gauge degrees of freedom behaves very
differently in a $2+1$ dimensional theory with a \cs term present for the gauge
field. There are three separate possibilities :
\begin{itemize}
\item The gauge masses are produced by the Higgs mechanism alone.
\item The gauge masses are produced by a \cs term alone.
\item The gauge masses are produced by both a Higgs potential and a \cs term.
\end{itemize}
It is sufficient to illustrate these cases with an abelian theory. The first
case corresponds to a Lagrange density
\barr
{\cal L}&=&-{1\over 4e^2} F_{\mu\nu}F^{\mu\nu} -\left(D_\mu \phi\right)^\dag
D^\mu \phi -V(\phi)
\label{higgs}
\earr
where $V(\phi)$ has some nontrivial vacuum $\phi_{(0)}$. Note that $e^2$ has
dimensions of mass in $2+1$ dimensions. In the broken vacuum, after shifting
the scalar field $\phi$ by $\phi_{(0)}$, we find the quadratic part of the
gauge field Lagrange density to be
\barr
{\cal L}_{quad}=-{1\over 4e^2} F_{\mu\nu}F^{\mu\nu}-|\phi_{(0)}|^2 A_\mu A^\mu
\label{higgsquad}
\earr
from which we deduce, as usual, a gauge mass
\barr
m_{gauge}=\sqrt{2} e |\phi_{(0)}|
\label{higgsmass}
\earr
However, if the Lagrange density includes also a \cs term, then the quadratic
part of the gauge Lagrange density in the broken vacuum is
\barr
{\cal L}_{quad}=-{1\over 4e^2} F_{\mu\nu}F^{\mu\nu}-{\kappa\over
2e^2}\epsilon^{\mu\nu\rho}A_\mu\partial_\nu A_\rho-|\phi_{(0)}|^2 A_\mu A^\mu
\label{mcshiggsquad}
\earr
which has {\it two} massive degrees of freedom
\barr
m_\pm={\kappa\over 2}\left( \sqrt{1+{8e^2|\phi_{(0)}|^2\over
\kappa^2}}\pm1\right)
\label{2masses}
\earr
Notice that with the \cs coupling in (\ref{mcshiggsquad}), the \cs coupling
parameter $\kappa$ has dimensions of mass. These two masses (\ref{2masses}) may
be deduced from the gauge propagator in a covariant gauge \cite{pisarski}, by a
self-dual factorization of the Maxwell-\cs-Proca equation \cite{paul}, or by a
Schr\"odinger representation analysis of the quadratic Hamiltonian
\cite{dunne7}.

The third possibility (the one that is realized in the \r \sd \cs systems
(\ref{rlag}) considered in this paper) is that the Lagrange density has a \cs
term but no Maxwell term \cite{deser3}. Then, in the broken vacuum, the
quadratic part of the gauge Lagrange density is
\barr
{\cal L}_{quad}=-{\kappa\over 2e^2}\epsilon^{\mu\nu\rho}A_\mu\partial_\nu
A_\rho-|\phi_{(0)}|^2 A_\mu A^\mu
\label{cshiggsquad}
\earr
from which we deduce a single massive mode, with mass
\barr
m_{CSH}={2e^2 |\phi_{(0)}|^2 \over\kappa}
\label{singlemass}
\earr
This pure \cs Higgs mechanism (\ref{cshiggsquad},\ref{singlemass}) can be
considered as the limit of the Maxwell-\cs Higgs mechanism
(\ref{mcshiggsquad},\ref{2masses}) in which $e^2\to\infty$ with $\kappa/e^2$
fixed.

A simple physical picture of these three different forms of gauge mass
generation in $2+1$ dimensions comes from the analogy of \cs quantum mechanics
\cite{dunne8}, in terms of which the conventional Higgs mechanism corresponds
to the planar quantum mechanics of a particle in a harmonic well, while the
Maxwell-\cs-Higgs mechanism corresponds to the planar quantum mechanics of a
particle in a harmonic well {\it and} a perpendicular external magnetic field.
In this latter case, there are two characteristic frequencies, and these are
precisely the two masses found in (\ref{2masses}) \cite{dunne7}. The pure \cs
Higgs mechanism corresponds to the lowest Landau level projection in which the
external magnetic field becomes very strong, so that the cyclotron frequency
scale is `frozen out' leaving a single frequency scale which matches the mass
in (\ref{singlemass}).

Having discussed the general situation, we now return to the specific case of
the \r \sd \cs system with Lagrange density (\ref{rlag}), regarded as a
symmetry breaking problem. The scalar masses in the vacuum $\phi_{(0)}$ are
determined by expanding the shifted potential $V(\phi+\phi_{(0)})$ to quadratic
order in the field $\phi$:
\barr
V(\phi +\phi_{(0)})&=&{v^4\over 4\kappa^2}tr\left( \left|
[[\phi_{(0)},\phi^\dag
],\phi_{(0)}] + [[\phi, \phi_{(0)}^\dag ], \phi_{(0)} ]+[[\phi_{(0)},
\phi_{(0)}^\dag ], \phi ] - \phi \right|^2  \right)\cr
\label{scalarmasses}
\earr
With the fields normalized appropriately, the masses are then given by the
square roots of the eigenvalues of the $2(N^2-1)\times 2(N^2-1)$ mass matrix in
(\ref{scalarmasses}).

In the unbroken vacuum, with $\phi_{(0)}=0$, there are $N^2-1$ complex
scalar fields, each with mass
\barr
m={v^2\over 2\kappa}
\label{scmass}
\earr
In one of the broken vacua, where $\phi_{(0)}\neq 0$, some of these $2(N^2-
1)$ massive scalar degrees of freedom are converted to massive gauge
degrees of freedom. The gauge masses are determined by expanding $v^2 tr\left(
\left( D_\mu\left(\phi+\phi_{(0)}\right)\right) ^\dag \left( D_\mu
\left(\phi+\phi_{(0)}\right)\right)\right)$ and extracting the piece quadratic
in the gauge field $A$:
\barr
v^2 \; tr\left( [A_\mu ,\phi_{(0)}]^\dag [A^\mu , \phi_{(0)} ]\right)
\label{gaugemasses}
\earr
The gauge masses are determined by finding the eigenvalues ({\it not} the
square roots of the eigenvalues) of the $(N^2-1)\times (N^2-1)$ mass matrix in
(\ref{gaugemasses}). Allowing for the nonabelian normalization factors, we see
from (\ref{singlemass}) that the gauge masses are given by multiples of the
{\it same} mass scale, $v^2/2\kappa$, as the scalar masses.

This procedure of finding the eigenvalues of the scalar and gauge mass
matrices, must be performed for each of the $p(N)$ gauge inequivalent
minima $\phi_{(0)}$ of $V$. The results for $SU(3)$ and $SU(4)$ are
presented here in Tables \ref{su3} and \ref{su4} (see also Ref. \cite{dunne6}).

\begin{table}[h]
\center
\begin{tabular}{|c|cc|cccc|} \hline
\multicolumn{1}{|c|}{\rm vacuum }&\multicolumn{6}{c|}{\rm gauge
masses}\\ \cline{2-7}
\multicolumn{1}{|c|}{\rm $\phi_{(0)}$}&\multicolumn{2}{c|} {\rm real}&
\multicolumn{4}{c|}{\rm complex}\\
\multicolumn{1}{|c|}{\rm }&\multicolumn{2}{c|}{\rm fields}&
\multicolumn{4}{c|}{\rm fields}\\ \hline\hline
$o-\times$&2&&1/2&1/2&1& \\ \hline
$o-o$&2&6&1&2&5& \\ \hline
\multicolumn{7}{c}{\rm } \\ \hline
\multicolumn{1}{|c|}{\rm vacuum }&\multicolumn{6}{c|}{\rm scalar
masses}\\ \cline{2-7}
\multicolumn{1}{|c|}{\rm $\phi_{(0)}$}&\multicolumn{2}{|c|} {\rm real}&
\multicolumn{4}{c|}{\rm complex}\\
\multicolumn{1}{|c|}{\rm }&\multicolumn{2}{c|}{\rm fields}&
\multicolumn{4}{c|}{\rm fields}\\ \hline\hline
$o-\times$&2&&1&3/2&3/2&2 \\ \hline
$o-o$&2&6&2&3&5& \\ \hline
\end{tabular} \\
\caption{$SU(3)$ vacuum mass spectra, in units of the fundamental mass scale
${v^2\over 2\kappa}$, for the inequivalent nontrivial minima $\phi_{(0)}$ of
the potential $V$. Notice that for each vacuum the {\it total} number of
massive degrees of freedom is equal to $2(N^2-1)=16$, although the distribution
between gauge and scalar fields is vacuum dependent.}
\label{su3}
\end{table}

A number of interesting observations can be made at this point, based on the
evaluation of these mass spectra for the various vacua in $SU(N)$ for $N$ up to
$10$.

(i) All masses, both gauge and scalar, are integer or half-odd-integer
multiples of the fundamental mass scale $m=v^2/2\kappa$. The fact that all
the scalar masses are proportional to $m$ is clear from the form of the
potential $V$ in (\ref{rpot}). The fact that the gauge masses are multiples of
the {\it same} mass scale depends on the fact that the Chern-Simons coupling
parameter $\kappa$ has been included in the overall normalization of the
potential in (\ref{rpot}). This is a direct consequence of the self-duality of
the model.

(ii) In each vacuum, the masses of the real scalar excitations are equal
to the masses of the real gauge excitations, whereas this is not true of the
complex scalar and gauge fields (by `complex' gauge fields we simply
mean those fields which naturally appear as complex combinations of the
nonhermitean step operator generators). Indeed, in some vacua the {\it
number} of complex scalar degrees of freedom and complex gauge degrees of
freedom is not even the same. This will be discussed further below.

(iii) In each vacuum, each mass appears at least twice, and always an
even number of times. For the complex fields this is a triviality, but for the
real fields this is only true as a consequence of the feature mentioned in
(ii). This pairing of the masses is a reflection of the $N=2$ supersymmetry
of the relativistic self-dual Chern-Simons systems \cite{clee,ivanov}.

(iv) While the distribution of masses between gauge and scalar modes is
different in the different vacua, the total number of degrees of freedom is, in
each case, equal to $2(N^2-1)$, as in the unbroken phase.

\begin{table}[h]
\center
\begin{tabular}{|c|ccc|ccccccccc|} \hline
\multicolumn{1}{|c|}{\rm vacuum }&\multicolumn{12}{c|}{\rm gauge
masses}\\ \cline{2-13}
\multicolumn{1}{|c|}{\rm $\phi_{(0)}$}&\multicolumn{3}{|c|} {\rm real}&
\multicolumn{9}{c|}{\rm complex}\\
\multicolumn{1}{|c|}{\rm }&\multicolumn{3}{c|}{\rm fields}&
\multicolumn{9}{c|}{\rm fields}\\ \hline\hline
$o-\times -\times$&2&&&1/2&1/2&1/2&1/2&1&&&& \\ \hline
$o-\times -o$&2&2&&1&1&1&1&2&&&& \\ \hline
$o-o-\times$&2&6&&1&1&1&2&2&5&&& \\ \hline
$o-o-o$&2&6&12&1&2&3&5&8&11&&& \\ \hline
\multicolumn{7}{c}{\rm } \\ \hline
\multicolumn{1}{|c|}{\rm vacuum }&\multicolumn{12}{c|}{\rm scalar
masses}\\ \cline{2-13}
\multicolumn{1}{|c|}{\rm $\phi_{(0)}$}&\multicolumn{3}{|c|} {\rm real}&
\multicolumn{9}{c|}{\rm complex}\\
\multicolumn{1}{|c|}{\rm }&\multicolumn{3}{c|}{\rm fields}&
\multicolumn{9}{c|}{\rm fields}\\ \hline\hline
$o-\times-\times$&2&&&1&1&1&1&3/2&3/2&3/2&3/2&2 \\ \hline
$o-\times-o$&2&2&&1&1&1&2&2&2&2&2& \\ \hline
$o-o-\times$&2&6&&1&2&2&2&2&3&5&& \\ \hline
$o-o-o$&2&6&12&2&3&4&5&8&11&&& \\
\hline
\end{tabular} \\
\caption{$SU(4)$ vacuum mass spectra, in units of the fundamental mass scale
${v^2\over 2\kappa}$, for the inequivalent nontrivial minima $\phi_{(0)}$ of
the potential $V$. Notice that for each vacuum the {\it total} number of
massive degrees of freedom is equal to $2(N^2-1)=30$, although the distribution
between gauge and scalar fields is vacuum dependent.}
\label{su4}
\end{table}

The most complicated, and most interesting, of the nontrivial vacua is the
``maximal embedding'' case, with $\phi_{(0)}$ given by (\ref{max}). For this
vacuum, the gauge and scalar mass spectra have additional features of note.
First, this ``maximal embedding'' also corresponds to ``maximal symmetry
breaking'', in the sense that in this vacuum all $N^2-1$ gauge degrees of
freedom acquire a mass. The original $2(N^2-1)$ massive scalar modes
divide equally between the scalar and gauge fields. The mass spectrum reveals
an intriguing and intricate pattern, as shown in Table \ref{sun}. It is
interesting to note that for the $SU(N)$ maximal symmetry breaking vacuum, the
entire scalar mass spectrum is {\it almost} degenerate with the gauge mass
spectrum : there is just {\it one} single complex component for which the
masses differ!

\begin{table}[t]
\center
\begin{tabular}{|c|ccccccc|} \hline
\multicolumn{8}{|c|}{\rm gauge masses}\\ \hline
\multicolumn{1}{|c|}{\rm real}&\multicolumn{7}{c|}{\rm complex} \\
\multicolumn{1}{|c|}{\rm fields}&\multicolumn{7}{c|}{\rm fields}\\
\hline
2&1&2&3&4&5&\dots&N-1  \\
6&5&8&11&14&\dots&3N-4& \\
12&11&16&21&\dots&5N-9&&\\
20&19&26&\dots&7N-16&&&\\
30&29&\dots&9N-25&&&& \\
\vdots&\vdots&&&&&&\\
N(N-1)&N(N-1)-1&&&&&&\\ \hline
\multicolumn{8}{c}{\rm } \\ \hline
\multicolumn{8}{|c|}{\rm scalar masses} \\ \hline
\multicolumn{1}{|c|}{\rm real}&\multicolumn{7}{c|}{\rm complex}\\
\multicolumn{1}{|c|}{\rm fields}&\multicolumn{7}{c|}{\rm fields}\\
\hline
2&N&2&3&4&5&\dots&N-1 \\
6&5&8&11&14&\dots&3N-4& \\
12&11&16&21&\dots&5N-9&&   \\
20&19&26&\dots&7N-16&&& \\
30&29&\dots&9N-25&&&& \\
\vdots&\vdots&&&&&& \\
N(N-1)&N(N-1)-1&&&&&&  \\ \hline
\end{tabular} \\
\caption{$SU(N)$ mass spectrum, in units of the fundamental mass scale
${v^2\over 2\kappa}$, for the maximal symmetry breaking vacuum, for which
$\phi_{(0)}$ is given by (\protect{\ref{max}}). Notice that the gauge mass
spectrum and the scalar mass are {\it almost} degenerate - they differ in just
one complex field component.}
\label{sun}
\end{table}

\b

\n{3.4 {\it {Mass Matrices for Real Fields}}

\b

The masses of the real fields exhibit further special simple properties, which
we discuss in this section. As mentioned above, in each vacuum $\phi_{(0)}$ the
{\it number} of real scalar modes is equal to the number of real gauge modes.
Furthermore, the two mass spectra coincide exactly, and are all {\it integer}
multiples of the mass scale $m$ in (\ref{mass}). The real gauge fields come
from the diagonal algebraic components $H_a$, while the real scalar fields come
from the simple root step operator components $E_a$. Indeed, the real scalar
fields correspond to those fields shifted by the symmetry breaking minimum
field $\phi_{(0)}$, which is decomposed in terms of the simple root step
operators as in (\ref{expansion}). This means that the {\it number} of real
scalars in a given vacuum $\phi_{(0)}$ is given by the number of nonzero
coefficients $\phi_{(0)}^a$ in the decomposition (\ref{expansion}). This can be
seen explicitly for $SU(3)$ and $SU(4)$ in the Tables \ref{su3} and \ref{su4}.
This also serves as an easy count of the number of real gauge masses. This also
means that to determine the mass matrix for the real gauge fields we can expand
$A_\mu$ in terms of the Cartan subalgebra elements $H_a$ (the other,
off-diagonal, algebraic components do not mix with these ones at quadratic
order). In fact, in order to normalize the gauge fields correctly, it is more
convenient to expand the $A_\mu$ in another Cartan subalgebra basis, $h_a$, for
which the traces are orthonormal (in contrast to the traces (\ref{chevalley})
in the Chevalley basis which involve the Cartan matrix) :
\barr
tr \left( h_a h_b \right)= \delta_{a\, b}
\label{newtrace}
\earr
Such basis elements, $h_a$, are related to the Chevalley basis elements, $H_a$,
by
\barr
h_a = \sum_{b=1}^r \omega_a^{(b)}\; H_b
\earr
where $\vec{\omega}^{(b)}$ is the $b^{th}$ fundamental weight of the algebra
\cite{humphreys}, satisfying
\barr
\sum_{b=1}^r \omega_a^{(b)} \alpha_c^{(b)} = \delta_{ac}
\label{orthog}
\earr
where $\vec{\alpha}^{(b)}$ is the $b^{th}$ simple root. For $SU(N)$ we can be
more explicit:
\barr
h_a = {1\over \sqrt{a(a+1)}}\sum_{b=1}^a b\, H_b
\earr
The orthogonality relation (\ref{orthog}) means that the correspondence can be
inverted to give
\barr
H_a = \sum_{b=1}^r \alpha_b^{(a)} \; h_b
\earr
The fundamental weights $\vec{\omega}^{(b)}$ and simple roots
$\vec{\alpha}^{(b)}$ are also related by
\barr
\vec{\alpha}^{(a)} = \sum_{b=1}^r C_{b a} \; \vec{\omega}^{(b)}
\earr
These new basis elements have the following commutation relations with the
simple root step operators:
\barr
[ h_a, E_b ] = \alpha_a^{(b)} E_b
\label{newcomm}
\earr
Given the traces in (\ref{newtrace}) and the commutation relations
(\ref{newcomm}), it is now a simple matter to expand the quadratic gauge field
term (\ref{gaugemasses}) to find the following mass matrix:
\barr
{\cal M}_{a b}^{\rm (gauge)} =  2\;m\;\sum_{c=1}^r |\phi_{(0)}^c |^2\,
\alpha_a^{(c)}\,\alpha_b^{(c)}
\label{gaugemass}
\earr
where $m$ is the fundamental mass scale in (\ref{mass}). For the maximal
embedding vacuum (\ref{max}) in $SU(N)$ this leads to a mass matrix
\barr
{\cal M}_{a b}^{\rm (gauge)} =  m\;\sum_{c=1}^{N-1} c(N-c)\,
\alpha_a^{(c)}\,\alpha_b^{(c)}
\label{sungaugemass}
\earr
This matrix has eigenvalues
\barr
2,\;6,\;12,\;20,\; \dots\;,\:N(N-1)
\label{eigenvalues}
\earr
in multiples of $m$. For any vacuum $\phi_{(0)}$ other than the maximal
symmetry breaking one, the mass matrix for the real gauge fields decomposes
into smaller matrices of the same form, according to the particular partition
of the original $SU(N)$ Dynkin diagram, as described in Section 3.2.

The real scalar field mass matrix can be computed by expanding the $\phi$ field
appearing in (\ref{scalarmasses}) in terms of the positive root step operators.
With such a decomposition for $\phi$, the quadratic term (\ref{scalarmasses})
simplifies considerably to give a mass (squared) matrix
\barr
{\cal M}_{a b}^{\rm (scalar)}= 4\;m^2\; \phi_{(0)}^a \phi_{(0)}^b \sum_{c=1}^r
|\phi_{(0)}^c |^2 \, C_{a c}\, C_{b c}
\earr
where $C$ is the Cartan matrix (\ref{cartan}). For the $SU(N)$ maximal symmetry
breaking vacuum (\ref{max}) this mass matrix is
\barr
{\cal M}_{a b}^{\rm (scalar)}= m^2\;
\sqrt{a\,b\,(N-a)\,(N-b)}\;\sum_{c=1}^{N-1} c\,(N-c)\, C_{a c}\, C_{b c}
\earr
which has eigenvalues
\barr
(2)^2,\;(6)^2,\;(12)^2,\;(20)^2,\;\dots\;,\;(N(N-1))^2
\label{eigs}
\earr
in units of $m^2$. It is interesting to note that the eigenvalues in
(\ref{eigs}) are the squares of the eigenvalues (\ref{eigenvalues}) of ${\cal
M}^{\rm (gauge)}$, even though ${\cal M}^{\rm (scalar)}$ is {\it not} the
square of the matrix ${\cal M}^{\rm (gauge)}$ in this basis. Nevertheless, as
the real scalar masses are given by the square roots of the eigenvalues in
(\ref{eigs}), we see that the real scalar masses do indeed coincide with the
real gauge masses, a consequence of the $N=2$ supersymmetry of the theory.

\section{Conclusion}

In these lectures I have reviewed certain selected aspects of \sd \cs theories.
The \sd \cs theories are $2+1$ dimensional models of charged scalar fields
interacting with gauge fields whose dynamics is described by a \cs Lagrangian
rather than a Maxwell-Yang-Mills Lagrangian. Both \nr and \r dynamics for the
scalar fields may be considered, and in each case there exists a classical
notion of self-duality whereby the classical energy functional is minimized by
solutions of first-order self-duality equations. In the \nr case, the \sd
equations are integrable and we have a complete understanding of the static \sd
solutions. In the \r case, the abelian \sd equations have been shown to fail a
Painlev\'e test for integrability, as does the $2+1$ dimensional abelian Higgs
model \cite{schiff}. Nevertheless, the existence of vortex-like solutions has
been established for the \sd \cs system \cite{wang}, just as for the abelian
Higgs model \cite{jaffe}. Even though general exact solutions are not
available, many properties of these \cs solitons may be deduced from asymptotic
and/or numerical information \cite{jackiw4,klee1}. The vacuum structure of
these \sd theories exhibits a rich structure, on which perturbative analyses of
the quantum theory will be based. The only known exact classical solutions are
gauge transforms of the constant fields which minimize the potential
(\ref{rpot}), which saturate the \sd energy bound. It would be interesting to
explore the possibility of finding other, less trivial, solutions - possibly by
some restrictive algebraic ansatze and/or by restricting to radially symmetric
solutions \cite{schiff}. It is also important to search for time dependent
solutions. In the \nr case one can generate time dependent solutions by
transforming static solutions, using the dynamical conformal symmetry of the
system \cite{jackiw2}. However, nothing is known about other truly time
dependent finite energy non-\sd solutions.

At the classical level of the Lagrangian and the equations of motion the \sd
\cs systems exhibit a rich space-time symmetry structure. The self-dual sixth
order potential (\ref{rpot}) in the \r \sd \cs theory may be fixed by requiring
that the Lagrange density (\ref{rlag}) be embedded into a supersymmetric theory
with an extended $N=2$ supersymmetry \cite{clee,ivanov}. This is consistent
with the general relationship between self-duality and extended supersymmetry
\cite{kao2,hlousek}. A similar property holds for the \nr \sd \cs system.
There, the fourth order form of the \sd potential in (\ref{nrlag}) may be fixed
by requiring that the Lagrange density (\ref{nrlag}) be embedded into a theory
with an $N=2$ superconformal Galilean symmetry \cite{martin1}. This entire
picture may be generalized to include {\it both} \cs {\it and} Maxwell dynamics
for the gauge field, in which case the gauge field is truly dynamical. Such an
extension requires the inclusion of additional scalar fields, in both the \r
\cite{leeleemin} and \nr \cite{dt,martin1} cases. These extra fields may be
interpreted as extra superpartners in a model with extended supersymmetry.

The most interesting open questions concern the {\it quantization} of the \sd
\cs theories. For the \nr \sd \cs system, the quantized field theory is a \nr
quantum field theory whose multi-particle sector corresponds to the
multi-particle quantum mechanics of anyons, and which provides a field
theoretic description of Aharonov-Bohm scattering \cite{bak1}. The \r \sd \cs
system is a quantum field theory of anyons. One can then ask: what is the {\it
quantum} significance of the {\it classical} self-duality symmetry which
minimizes the classical energy functional? In the \nr system perturbative
analyses of Aharonov-Bohm scattering indicate that the quartic potential, which
corresponds quantum mechanically to a $\delta$-function hard-core
inter-particle potential, is necessary for renormalization \cite{giovanni};
and, moreover, the classical conformal invariance is preserved at the \sd point
\cite{bak2}. It has also been shown that the one-loop contribution to the
effective potential vanishes with the \sd quartic self-interaction
\cite{gustavo2}. Considerably less is known about the quantization of the \r
\sd \cs systems. One would like to understand better the quantum significance
of the classical \sd solitons in, for example, a collective coordinate
formulation. Further issues, such as renormalization \cite{klee3}, vacuum
tunnelling, and the perturbative fate of the \sd potential remain to be
resolved.

\b

\section{Acknowledgement :} I would like to thank Professors Jihn E. Kim and
Choonkyu Lee, and all the participants, for an extremely interesting and
enjoyable symposium.

\end{document}